\documentclass[reprint, twocolumn, notitlepage,nofootinbib]{revtex4-2}
\usepackage{graphicx}
\usepackage{xcolor}
\usepackage{amsmath, amssymb, amsfonts, amsthm, bm, dsfont, braket, physics, upgreek} 
\usepackage[colorlinks=true, allcolors=blue]{hyperref}
\usepackage[capitalise]{cleveref}
\crefformat{section}{\S#2#1#3}

\newcommand{\Integers}{\ensuremath{\mathds{Z}}}
\newcommand{\Reals}{\ensuremath{\mathds{R}}}

\newcommand{\T}{\ensuremath{{\scriptstyle\mathsf{T}}}}

\begin{document}
\title{Quantum-Assisted Support Vector Regression}
\author{Archismita Dalal}
\thanks{Corresponding author}
\email{archismita.dalal1@ucalgary.ca}
\author{Mohsen Bagherimehrab}
\author{Barry C.\ Sanders}
\affiliation{Institute for Quantum Science and Technology, University of Calgary, Alberta T2N~1N4, Canada}
\date{\today}

\begin{abstract}
A popular machine-learning model for regression tasks, including stock-market prediction, weather forecasting and real-estate pricing, is the classical support vector regression~(SVR).
However, a practically realisable quantum SVR remains to be formulated.
We devise annealing-based algorithms, namely simulated and quantum-classical hybrid, for training two SVR models, and compare their empirical performances against the SVR implementation of Python's scikit-learn package for facial landmark detection (FLD), a particular use case for SVR. 
Our method is to derive a quadratic-unconstrained-binary formulation for the optimisation problem used for training a SVR model and solve this problem using annealing. 
Using D-Wave’s Hybrid Solver, we construct a quantum-assisted SVR model, thereby demonstrating a slight advantage over classical models regarding FLD accuracy.
Furthermore, we observe that annealing-based SVR models predict landmarks with lower variances compared to the SVR models trained by gradient-based methods.
Our work is a proof-of-concept example for applying quantum-assisted SVR to a supervised learning task with a small training dataset.
\end{abstract}
\maketitle

\makeatletter
\renewcommand*\l@section{\@dottedtocline{1}{0em}{1em}}
\renewcommand*\l@subsection{\@dottedtocline{2}{1em}{1.5em}}
\renewcommand*\l@subsubsection{\@dottedtocline{3}{2.5em}{2.5em}}
\makeatother

\section{Introduction}
The classical machine-learning model for support vector regression~(SVR) is widely used for regression tasks, including prediction of weather, stock market and real-estate pricing~\cite{DBK+97, SS04, RS09, HSK18, LXZR09}.
However, a practically realisable quantum version for SVR is yet to be established in the quantum machine-learning domain~\cite{BWP+17, DB18}.
Current feasible applications of quantum machine learning employ quantum annealing~\cite{KN98} to enhance one of the essential components in machine learning, i.e., optimisation problem.
Consequently, these quantum-assisted solutions are shown, empirically, to be slightly more accurate than classical solutions~\cite{MJV+17, LFRL18, WWDM20}.
Our aim here is to devise a quantum-assisted SVR model by employing D-Wave's quantum-classical hybrid solvers~\cite{leaphybrid},
and compare this quantum-assisted model against classical models for detecting facial landmarks, e.g.,
centres of eyes, nose tip and corners of the mouth,
in unconstrained images~\cite{WGT+18}.

We state the facial landmark detection~(FLD) task, along with its applications, computational challenges and state-of-the-art methods to accurately perform this task.  
The task of FLD is to identify key landmarks on a human-face image~\cite{WGT+18, WJ19},
with important applications such as face recognition~\cite{MCC+18,ZLY+15,TYRW14}, three-dimensional face reconstruction~\cite{CML+10}, facial-emotion recognition~\cite{NTP+17} and gender prediction\cite{RPC17}.
Efficient and robust FLD is challenging due to large variability of appearance, expression, illumination and partial occlusion of unconstrained face images, i.e., images obtained in uncontrolled conditions~\cite{WGT+18}.
Neural regression-based algorithms are considered the state-of-the-art in FLD as they deliver the highest detection accuracies so far~\cite{SWT13,KK21}.

We introduce a quantum-assisted model to enhance the regression performance of SVR and further utilise FLD as a proof-of-principle application.
The detection accuracy of a FLD algorithm is typically limited by the size and quality of training data and computational resources available~\cite{KK21}.
To this end, we test if quantum resources can enhance the performance of our SVR-based FLD algorithm for the case of a small training dataset comprising~100 unconstrained face images.
We use a limited-size training dataset for two reasons: feasibility on current quantum hardwares and the success of other quantum machine-learning applications~\cite{MJV+17, LFRL18, WWDM20} with limited-size datasets.
Our FLD algorithm only serves as a test case for our quantum-assisted SVR formulation and does not directly advance the state-of-the-art in FLD algorithms~\cite{KK21}. 

Quantum-assisted algorithms are exhibited as superior alternatives to classical algorithms for classification tasks, including the protein-binding problem in computational biology~\cite{WWDM20, LFRL18} and the Higgs particle-classification problem in high-energy physics~\cite{MJV+17}. 
One such promising classification model is a quantum-assisted support vector machine~(SVM), which is mathematically similar to SVR and makes use of support vectors to train a classification model~\cite{SS18}.
The support vectors are calculated by solving a constrained optimisation problem using quantum annealing~\cite{WWDM20}, as opposed to using gradient-based routines in \texttt{scikit-learn}~\cite{sgd}.
D-Wave Systems' practical quantum annealing machine, commonly called an annealer, solves this optimisation problem by casting it as an Ising minimisation, or equivalently, quadratic unconstrained binary optimisation~(QUBO) problem.
The classification accuracy of a SVM model, trained with a limited-size training dataset, for the task of protein binding is improved using an ensemble of close-to-optimal solutions obtained from D-Wave's quantum annealer~\cite{WWDM20}.

Whereas D-Wave's quantum annealer has many applications, there are some limitations on its implementation of quantum annealing and its broader usability.
The process of sampling from the annealer is not deterministic; hence, the solution suffers from finite-sampling error. Moreover, physical noise sources in a quantum annealer further degrades the solution quality.
Although the state-of-the-art quantum annealer comprises about 5000 qubits, this device can solve optimisation problems with up to 180 variables, which correspond to fully connected graphs, due to restricted connectivity in its hardware.
As empirical evidence for quantum speedup with D-Wave's annealer is still under investigation, machine learning tasks are assessed using other performance metrics~\cite{LFRL18,WWDM20}.
Moreover, recent applications employ D-Wave's quantum-classical hybrid annealing, which can tackle problems with a million variables~\cite{HJK+20, PSH21}.

We develop a quantum-classical hybrid machine-learning algorithm to solve the FLD problem and execute this algorithm on a D-Wave's quantum annealer.
Methodologically, we split the multi-output regression task for FLD~\cite{BVBL15} into several single-output regressions and then train a SVR model for each single-output regression problem.
We derive a QUBO formulation for the constrained optimisation problem associated with the training of each SVR model and solve this QUBO problem using both classical annealing and hybrid approaches.
Having constructed the quantum-assisted and classical models, we then assess and compare the statistical significance of their detection accuracies for FLD.
We observe that the annealing-based SVR models predict landmarks with lower variance compared to the SVR models trained using gradient-based optimisation methods~\cite{sklearnSVR}.
Thus, in addition to introducing a new use case for quantum annealers and other quantum optimisation algorithms, our work also establishes the slight advantage of D-Wave's Hybrid Solver over simulated annealing.

Our paper is organised as follows.
We begin in~\cref{sec:background} by elaborating the pertinent background for SVR, FLD and quantum annealing on D-Wave. In~\cref{sec:approach}, we describe our approach for solving the FLD problem using quantum-assisted SVR models.
We then present our results in~\cref{sec:results}, where we derive a QUBO formulation for SVR and compare the performance of our quantum-assisted approach against the classical approaches for FLD.
Finally, we discuss our results and their implications in~\cref{sec:discussion}, and conclude in~\cref{sec:conclusion}.

\section{Background}
\label{sec:background}
In this section, we discuss the relevant background for SVR machines, regression-based FLD and practical quantum annealing.
We begin, in~\cref{subsec:svr}, by explaining the supervised-learning problem of regression and the SVR model.
Then, in~\cref{subsec:fld}, we present the FLD problem as a supervised learning problem of regression, along with its standard performance metrics.
In the final subsection~\cref{subsec:D-Wave}, we describe the concepts of quantum annealing and its implementation by D-Wave Systems.

\subsection{Support vector regression}
\label{subsec:svr}
We now review the key background pertinent to SVR~\cite{DBK+97, SS04}, which is a tool for solving the supervised-learning~(SL) problem of regression.
A SVR has two formulations known as ``primal'' and ``dual''; we begin by describing the dual formulation of a linear SVR, with the background on primal formulation provided in~\cref{appendix:svr_basics}.
Then we obtain an expression for the linear SVR model, i.e., the prediction function, in terms of the variables in the dual formulation.
Finally, we discuss the kernel method that is used to deal with nonlinear regression and state the commonly used kernel functions.

For a SL problem of single-output regression, we are given a training dataset with $M$ data points.
Each data point is a tuple $(\bm{x}_i,y_i)$, where $\bm{x}_i$ is a real-valued feature vector of dimension~$F$ and $y_i$ is the target value of~$\bm{x}_i$.
The dataset is formally stated as 
\begin{equation}
\label{eq:train_data}
  \mathcal{D}_{\text{SVR}}=\{(\bm{x}_i,y_i) \mid i\in [M]:=\{0,\ldots, M-1\}\} \subset \Reals^F \times \Reals.
\end{equation}
The learning problem involves estimating a prediction function~$f(\bm{x})$ for an unseen feature vector~$\bm{x}$, such that the estimated target is~$\tilde y=f(\bm{x})$.
SVR is a powerful and robust method to address this learning problem~\cite{DBK+97, SS04}.
For~$\bm w$ the normal vector to a hyperplane and~$b$ the offset,
the task in the linear SVR is to search for a linear prediction function
\begin{equation}
\label{eq:linear_function}
    f_\text{linear}(\bm{x})=\bm{w} \cdot \bm{x}+b \quad (\bm{w} \in \Reals^F, b \in \Reals),
\end{equation}
such that, for a given error tolerance~$\varepsilon \in \Reals^+$, 
\begin{equation}
\label{eq:constrains}
  \abs{f_\text{linear}(\bm{x}_i)-y_i} \leq \varepsilon \quad \forall i\in [M],
\end{equation}
while minimizing the norm $\norm{\bm w}^2=\bm w \cdot \bm w$~\cite{SS04}.
This formulation is also known as~$\varepsilon$-SVR, where $\varepsilon$ is the error tolerance in~\cref{eq:constrains}.

In the dual formulation of SVR, we construct a Lagrange function from the objective function~\eqref{eq:primal} and the constraints~\eqref{eq:primalconstraints} of the primal optimisation problem using Lagrange multiplier vectors $\bm{\alpha}^+, \bm{\alpha}^- \in (\Reals^+)^M$.
Lagrange equations are obtained by taking partial derivatives of Lagrange function with respect to $\bm{w}$, $b$ and the two slack variables~(\cref{appendix:svr_basics}), and then setting these equations to zero.
This procedure yields an expression for $\bm{w}$ in terms of $\bm{\alpha}^+, \bm{\alpha}^-$ as
\begin{equation}
\label{eq:normal_vector}
    \bm{w} =\left(\bm{\alpha}^+-\bm{\alpha}^-\right)\cdot \bm X,
\end{equation}
where~$\bm X$ is a vector that comprises the feature vectors~$\{\bm{x}_i\}$ in~\cref{eq:train_data} as its components, and constraints
\begin{equation}
\label{eq:dualconstrains}
     \norm{\bm\alpha^+}_1= \norm{\bm\alpha^-}_1,\; 0\leq \alpha^+_i, \alpha^-_i\leq \gamma
    \quad \forall m \in [M],
\end{equation}
on~$\bm{\alpha}^+$ and $ \bm{\alpha}^-$.
The optimisation problem in the dual formulation of $\varepsilon$-SVR is
\begin{align}
    \label{eq:dual_optimisation}
    \min_{\bm{\alpha}^+, \bm{\alpha}^-}
    \Bigg\{
    \frac{1}{2}
    &(\bm\alpha^+ -\bm\alpha^-)^\T\bm{K}
    (\bm\alpha^+ -\bm\alpha^-)
    +\varepsilon \norm{\bm\alpha^+ +\bm\alpha^-}_1 \nonumber\\
    &-\bm{y}\cdot(\bm\alpha^+ -\bm\alpha^-)\Bigg\},
\end{align}
for
\begin{equation}
\bm K =(K_{ij}:=\bm{x}_i\cdot\bm{x}_j) \in \Reals^{M\times M},
\end{equation}
subject to the constraints in~\cref{eq:dualconstrains}.
The solution of this optimisation problem is the two Lagrange multiplier vectors~$\bm{\alpha}^+$ and $\bm{\alpha}^-$.
Only a few of the Lagrange multipliers are nonzero: these are known as support vectors and are the ones corresponding to data points that specify the prediction function.


Now we explain the standard method to convert the two-variable ($\bm{\alpha}^+$ and $ \bm{\alpha}^-$) optimisation problem~\eqref{eq:dual_optimisation} into a single-variable problem~\cite{DBK+97}.
This is done by defining a new Lagrange multiplier vector~$\bm{\alpha}$, whose elements relate to the two original Lagrange multiplier vectors as
\begin{equation}
\label{eq:new_multipliers}
   \alpha_i:= \alpha^+_i,\quad
\alpha_{M+i}:=\alpha^-_i \quad
\forall i \in [M].
\end{equation}
Additionally, $\bm{y}$ and $\varepsilon$ are replaced by a single new variable~$\bm{c}$ as
\begin{equation}
   c_i:= \varepsilon-y_i,\quad
c_{M+i}:=\varepsilon+y_i
\quad \forall i \in [M],
\end{equation}
and a $2M\times 2M$ symmetric matrix~$\bm{Q}$ is introduced as
\begin{equation}
    \bm{Q}:=\begin{bmatrix}
\bm{K} & -\bm{K}\\
-\bm{K} & \bm{K}
\end{bmatrix},
\end{equation}
with $\bm{K}$ the kernel matrix in~\cref{eq:dual_optimisation}.
In terms of the new variables~$\bm\alpha$ and $\bm c$, the one-variable optimisation problem in the dual form is
\begin{subequations}
\label{eq:minimize}
\begin{align}
    &\min_{\bm{\alpha}} \left\{ \frac{1}{2}\bm{\alpha}^\T\bm{Q}\bm{\alpha} + \bm{\alpha}\cdot\bm{c}\right\} \label{eq:minimize-1}, \\
    &\sum_{i=0}^{M-1}\alpha_i=\sum_{i=M}^{2M-1}\alpha_i,\;
    \alpha_i\in[0,\gamma]
    \quad\forall i\in[2M]. 
    \label{eq:minimize-2}
\end{align}
\end{subequations}
Due to the quadratic nature of this optimisation problem, the solution to this problem is a unique Lagrange multiplier vector~$\bm{\alpha} \in \Reals^{2M}$.

The linear prediction function~\eqref{eq:linear_function} is now expressed in terms of the Lagrange multiplier vector~$\bm{\alpha}$ and offset~$b$ as
\begin{equation}
\label{eq:prediction_function}
    f_\text{linear}(\bm{x})=\sum_{i=0}^{M-1}
    (\alpha^+_i-\alpha^-_i)\,\bm{x}_i\cdot\bm{x}
    + b,
\end{equation} 
where $b$ is calculated using~$\bm{\alpha}$; 
see~\cref{appendix:offset} for details.
This function is commonly referred to as the linear $\varepsilon$-SVR model.
In the following, we explain the formulation of a nonlinear extension of $\varepsilon$-SVR using a method known as the kernel method; see Ref.~\cite[Sec.~7.4]{wit14} for details.


Feature mapping is an approach for applying the linear $\varepsilon$-SVR formulation to nonlinear regression problems.
In this approach, vectors of the input feature space are first embedded into a space of equal or higher-dimension by a feature map\footnote{Throughout this paper we use \texttt{typewriter} font for functions and packages.} defined as
\begin{equation}
\label{eq:embed}
    \texttt{embed}: \Reals^F \to \Reals^{F'}:
    \bm{x}_m \mapsto \texttt{embed}(\bm{x}_m)
    \quad \forall\, F'\geq F
\end{equation}
and then a linear $\varepsilon$-SVR model~\eqref{eq:prediction_function} is constructed using the embedded feature vectors.
To accommodate nonlinear data, the feature-mapping approach requires an explicitly defined
feature map~$\texttt{embed}$~\eqref{eq:embed} and becomes computationally inefficient as $F'$ increases.

Kernel method bypasses the embedding in feature mapping and provides a computationally efficient way to extend the linear $\varepsilon$-SVR to nonlinear data.
This approach relies on the observation that the optimisation problem~\eqref{eq:dual_optimisation} in the dual formulation and the prediction function~\eqref{eq:prediction_function} depend only on the dot product between the feature vectors.
Therefore, we only need to know the dot product between~the embedded vectors $\texttt{embed}(\bm{x}_m)$ rather than the feature map $\texttt{embed}$~\eqref{eq:embed} itself to construct a prediction model.
In kernel methods, the dot product between the embedded vectors is computed by a kernel function 
\begin{align}
    &K: \Reals^F \times \Reals^F \to \Reals,\nonumber\\
    &(\bm{x}_n, \bm{x}_m) \mapsto K(\bm{x}_n, \bm{x}_m):= \texttt{embed}(\bm{x}_n)\cdot \texttt{embed}(\bm{x}_m).
\end{align}
Thus the optimisation problem for the nonlinear $\varepsilon$-SVR becomes analogous to that for the linear $\varepsilon$-SVR, but with the dot product between feature vectors being replaced by the kernel function.
Consequently, the prediction function in the kernel method becomes
\begin{equation}
\label{eq:prediction_function_nl}
    f(\bm{x})= \sum_{i=0}^{M-1}(\alpha^+_i-\alpha^-_i) K(\bm{x}_i,\bm{x})+b,
\end{equation}
which is a nonlinear version of the linear prediction function in~\cref{eq:prediction_function}.

The three kernel functions that are commonly used in machine-learning literature are linear, polynomial and Gaussian kernels.
The linear kernel, defined as~$K_\text{L}(\bm{x}_n, \bm{x}_m):= \bm{x}_n\cdot\bm{x}_m,$
corresponds to trivial embedding~\eqref{eq:embed} and is used for linear dataset.
The polynomial kernel, for a degree-$d$ polynomial,
\begin{align}
    \label{eq:Kpoly}
    K_\text{P}(\bm{x}_n, \bm{x}_m) = (\bm{x}_n\cdot\bm{x}_m+c)^d
    \quad
    \forall c\in \Reals, d\in\Integers^+,
\end{align}
and the Gaussian kernel 
\begin{align}
    \label{eq:KGaussian}
    K_\text{G}(\bm{x}_n, \bm{x}_m)
        =\text{e}^{-\eta\abs{\bm{x}_n-\bm{x}_m}^2}
        \quad  \forall\eta\in \Reals^+,
\end{align}
are used for nonlinear datasets.
If no prior knowledge is available to determine the hyperparameter~$\eta$ for Gaussian kernel, its default value is chosen to be 
\begin{equation}
\label{eq:default_eta}
    \eta = 1/(F\sigma^2),
\end{equation}
where $F$ is the number of features and $\sigma$ is the standard deviation of the given dataset~\cite{sklearnSVR}.

\subsection{Facial-landmark detection}
\label{subsec:fld}
In this subsection, we explain the relevant background on FLD.
We commence by describing the FLD task and then discuss how to convert this task into a computational problem by introducing the concept of data preprocessing.
Next we cast FLD as a SL problem and elaborate on the learning workflow.
Finally, we discuss standard benchmarking datasets and commonly used performance measures for evaluating a FLD algorithm.

We begin by defining a `face shape', which we use to describe the FLD problem.
A face shape is a collection of $(x,y)$ coordinates of a number~$L$ of key landmarks on the face;
$L$ is typically a number between~5 to~100 depending on the application~\cite{WGT+18}.
For a truecolor face image~$\bm{I}^\text{raw}$ with $L$ landmarks, we represent its face shape~$\bm{s}^\text{raw}$ by the vector
\begin{equation}
\label{eq:shape_list}
    \bm{s}^\text{raw}=(s^{\text{raw}}_0, s^{\text{raw}}_1, \ldots, s^{\text{raw}}_{2L-1}),
\end{equation}
where $s_{2k}^\text{raw}$ and $s_{2k+1}^\text{raw}$ are real numbers denoting the manually determined values for $x$ and $y$ coordinates of the~$k$th landmark, respectively.
For a FLD problem, we are given a raw dataset 
\begin{equation}
\label{eq:rawdata}
  \mathcal{D}^\text{raw}:=\{(\bm{I}^\text{raw}_i,\bm{s}^\text{raw}_i) \mid i\in [N]\} \subset \Integers^{m\times n \times 3} \times \Reals^{2L},
\end{equation}
where each~$\bm{I}^\text{raw}_i$ is a truecolor face image, 
represented by an $m\cross n\cross 3$ array of integers in the range [0, 255] that defines red, green and blue color components for each pixel of the image\footnote{In \texttt{OpenCV}~\cite{imageprop}, $m$ and $n$ denotes number of rows and columns of pixels, respectively.}, and~$\bm{s}^\text{raw}_i$ is the manually determined face shape\footnote{In \texttt{OpenCV}, the coordinate system is an inverted Cartesian system with origin a the top-left corner. Each coordinate can be further bounded as $s_{i,2k}^\text{raw}\in[0,m]$ and $s_{i,2k+1}^\text{raw}\in[0,n]$} corresponding to~$\bm{I}^\text{raw}_i$.
The task in FLD is to devise a model
\begin{equation}
\label{eq:shape}
\texttt{shape}: \Integers^{m\times n \times 3} \to \Reals^{2L} : \bm{I}^\text{raw} \mapsto \bm{s}^\text{raw},
\end{equation}
to accurately predict the face shape~$\bm{s}^\text{raw}$ for an unmarked raw image~$\bm{I}^\text{raw}$.

The raw dataset of marked truecolor images is first preprocessed before being used to devise a model for FLD.
Preprocessing is important because working directly with the raw dataset makes the FLD task computationally more expensive.
Moreover, the dataset of unconstrained truecolor images may vary largely in facial region size, face orientation, or illumination.
We provide a detailed description of the prepossessing steps in~\cref{appendix:preprocess} and provide our specifications for these steps in \cref{appendix:workflow}.
In short, preprocessing of a raw image~$\bm{I}^\text{raw}$ comprises three sequential operations. 
First, \texttt{normalise} constructs a normalised image~$\bm{I}^\text{norm}$ from the raw image.
Next, \texttt{extract} maps $\bm{I}^\text{norm}$ into a $F_\text{norm}$-dimensional feature vector~$\bm{x}^\text{norm}$.
Finally, \texttt{select} converts $\bm{x}^\text{norm}$ to a low-dimensional feature vector~$\bm{x}$ of size $F$.
Additionally, \texttt{scale} computes the face shape~$\bm{s}$ of~$\bm{I}^\text{norm}$.

The resultant dataset for a FLD problem after preprocessing the raw dataset~\eqref{eq:rawdata} is 
\begin{equation}
\label{eq:processed_data}
  \mathcal{D}=\{(\bm{x}_i,\bm{s}_i) \mid i\in [N]\} \subset \Reals^F \times \Reals^{2L}.
\end{equation}
Using this dataset, a model 
\begin{equation}
\label{eq:detect}
    \texttt{detect:} \Reals^F \to \Reals^{2L}: \bm{x} \mapsto \bm{s}
\end{equation}
is devised that maps the $F$-dimensional feature vector $\bm{x}$ to the face shape $\bm{s}$ of a normalised face image.
The face shape of a raw face image is then achieved by applying $\texttt{rescale}$, which is defined as the inverse of \texttt{scale}~\eqref{eq:scale}, to $\bm{s}$.
For $\texttt{preprocess}:=\texttt{select} \circ \texttt{extract} \circ \texttt{normalise}$,
finding a face shape is therefore the composition
\begin{equation}
\label{eq:shape_compose}
    \texttt{shape}= \texttt{rescale}\circ \texttt{detect}\circ \texttt{preprocess},
\end{equation}
which is the computational problem denoting a FLD task.

The literature shows a plethora of methods used to solve the FLD problem~\cite{WGT+18}, with more recent and efficient algorithms developed by regression-based methods~\cite{VMBP10, SWT13}. 
These methods learn a regression model \texttt{detect} from the low-dimensional feature vector $\bm{x}$ to the face shape  $\bm{s}$ of a normalised face image.
Ref.~\cite{VMBP10} employs a method that combines SVR for local search and Markov random fields for global shape constraints, and yields fast and accurate detection of landmarks.
We next cast FLD as a regression problem and describe a typical machine-learning workflow for solving the regression problem.

The FLD problem is cast as a SL problem of multi-output regression~\cite{BVBL15}. 
Using the preprocessed dataset $\mathcal{D}$~\eqref{eq:processed_data}, the SL algorithm learns a model to accurately predict the face shape of a face image. 
This SL problem follows a typical machine-learning workflow involving four steps, namely preprocessing the dataset, calibration of model hyperparameters, training, and evaluation of the model.
For SL, the preprocessed dataset can be divided as
\begin{equation}
\label{eq:ModelTest}
    \mathcal{D}=\mathcal{D}_\text{model}\sqcup \mathcal{D}_\text{test},
\end{equation}
where the size of $\mathcal{D}_\text{model}$ is~$M$ and size of $\mathcal{D}_\text{test}$ is~$N-~M$.
The model dataset $\mathcal{D}_\text{model}$ is used for calibrating and training a model, whereas the test dataset $\mathcal{D}_\text{test}$ is used to test the model.

In the calibration step, the hyperparameters of the model can be tuned in an iterative way~\cite{Burman89}.
In this technique, a tuple of optimal hyperparameters of the model is obtained by searching over the hyperparameter space.
For each tuple of hyperparameters, a model is trained on a randomly-sampled subset~$\mathcal{D}_\text{train}$ of $\mathcal{D}_\text{model}$, and the model's performance is evaluated on the remaining dataset.
A mean performance, corresponding to each tuple of hyperparameters, is then calculated by repeating these two sub-steps for different~$\mathcal{D}_\text{train}$. 
After repeating this process of calculating mean performance for all possible tuples of hyperparameters, the calibration step returns the tuple yielding the best performance.

In the training step, the model dataset $\mathcal{D}_\text{model}$, along with the hyperparameters returned from the calibration step, are used to construct a machine-learning model
\begin{equation}
\label{eq:approxdetect}
    \widehat{\texttt{detect}}: \Reals^F \to \Reals^{2L}: \bm{x} \mapsto \tilde{\bm{s}},
\end{equation}
which approximates the ideal model \texttt{detect}~\eqref{eq:detect}, and yields an approximate shape $\tilde{\bm{s}}$ of an unseen face image.
The performance of this model is  assessed on the test dataset $\mathcal{D}_\text{test}$, which is unseen in the calibration and training steps.
Additionally, a k-fold cross validation on $\mathcal{D}_\text{model}$ can be used for assessing learned models~\cite{Burman89}.
For comparing quality of different models for FLD, their performances on the same test dataset are calculated using the standard metrics, as explained next.

The datasets used for testing a ML model for FLD include images of human faces along with manually labelled landmarks~\cite{WGT+18}.
These datasets are constructed either under controlled conditions (constrained) or under uncontrolled conditions (unconstrained).
Different datasets also vary in the total number of marked landmarks.
Commonly used unconstrained datasets for FLD are BioID~2001~\cite{JKF01}, Labeled Faces in the Wild 2007~(LFW)~\cite{HRBL07}, Labeled Face Parts in the Wild 2011~(LFPW)~\cite{BJKK11} and Helen~2012~\cite{LBL+12},
whereas constrained datasets include IMM database~\cite{NLSS04}
and PUT database~\cite{KFS08}.

Two performance measures are typically used to assess a FLD model's performance:
mean normalised detection error~(MNDE)
and failure rate~(FR)~\cite{SWT13}.
The detection error for each landmark is the Euclidean distance between the observed and the predicted coordinates.
This error is normalised to make the performance measure independent of the actual face size or the camera zoom~\cite[p.~4]{CUS13}.
Conventionally, the detection error is normalised by dividing with the inter-ocular distance, which is the Euclidean distance between the centre of the eyes~\cite{CC06}.
However, this normalisation is biased for profile faces for which the inter-ocular distance can be very small~\cite{ZR12, SWT13}.
An alternative approach for normalisation, which does not have the drawback of the conventional normalisation, is dividing the detection error by the width of the face bounding box~\cite{SWT13}.
For each landmark~$k$ with the true coordinates~$(x_k,y_k)$ and the corresponding predicted coordinated~$(\tilde{x}_k,\tilde{y}_k)$, the normalised detection error
\begin{equation}
\label{eq:DE}
    e_k:= \sqrt{(x_k-\tilde{x}_k)^2+(y_k-\tilde{y}_k)^2}/d_k
\end{equation}
where $d_k$ is the inter-ocular distance or the width of the face bounding box. 
The MNDE for each landmark $k$ 
over a dataset with $N$ images
is defined as the arithmetic mean of the normalised detection error for the landmark in each image of the dataset; that is 
\begin{equation}
\label{eq:MNDE}
    \text{MNDE}_k :=\sum_{i=1}^{N} \frac{e^i_k}{N},
\end{equation}
where $e^i_k$ is the normalised detection error for $k$th landmark of $i$th image.

To avoid biases of the MNDE, due to the variations in error normalisation, FR is also used as a measure for performance of a FLD model.
For FR, a pre-specified threshed value, denoted by~$e_\text{th}$, is required for the normalised detection error~\eqref{eq:DE}.
If the normalised detection error is greater than~$e_\text{th}$ then the detected landmark is considered as a failed detection. 
For each landmark~$k$, the FR is defined as
\begin{equation}
\label{eq:FR}
    \text{FR}_k:=\frac{\abs{\left\{i: e^i_k>e_\text{th}\right\}}}{N},
\end{equation}
which is the ratio of number of failed detection to the the total number of images~$N$; the term `rate' here refers to the ratio.
The commonly used threshold value for failed detection is $e_\text{th}=0.1$~\cite[p.~4]{CUS13}.
We use both MNDE and FR to gauge the accuracy of a FLD algorithm and to compare the performance of different FLD algorithms.

\subsection{Quantum annealing on D-Wave}
\label{subsec:D-Wave}
In this subsection, we briefly discuss the concept of quantum annealing, which employs quantum-mechanical effects to approximate the solution of an optimisation problem, and its practical realisation.
We start by defining the Ising minimisation problem and its relation to the computational problem of optimisation.
We then define quantum annealing and explain how this physical process results in finding the solution of a minimisation problem.
Next we state the equivalence between an Ising minimisation problem and a QUBO problem.
Finally, we discuss D-Wave's implementation of quantum annealing, with applications to SL.

The computational problem of finding the global minimum is equivalent to the physical problem of finding the ground state of an Ising spin system, which is a collection of pairwise-interacting spin-$1/2$ particles in an external magnetic field.
For a spin configuration $\{\sigma_i^\text Z\}$, $h_i$ and~$J_{ij}$ represent the strength of the magnetic field on particle~$i$ and the coupling strength between adjacent particles~$i$ and $j$, respectively.
The energy of this system is expressed by the Ising Hamiltonian
\begin{equation}
\label{eq:ising_hamiltonian}
    H_\text P
    :=\sum_{i} h_i \sigma_i^\text Z + \sum_{<i,j>} J_{ij} \sigma_i^\text Z \sigma_j^\text Z,
\end{equation}
where the subscript `P' denotes problem and the notation $<i,j>$ denotes adjacency between particles~$i$ and~$j$.
Here~$\sigma_i^\text Z\in\{\pm1\}$ for this classical $H_\text P$, whereas for a quantum particle, $\sigma_i^\text Z$ represents the Pauli Z-matrix operating on particle~$i$.
Given coefficients $\{h_i\}$ and $\{J_{ij}\}$, the Ising minimisation problem is to find~$\{\sigma_i^\text Z\}$ such that the system achieves the minimum or ground state energy.

Quantum annealing is a meta-heuristic optimisation procedure that aims to find the global minimum of a discrete optimisation problem using properties of quantum physics~\cite{KN98}.
This optimisation problem is represented as an Ising minimisation problem by expressing the coefficients of the objective function in terms of $\{h_i\}$ and~$\{J_{ij}\}$, and mapping the discrete variables to $\{\sigma_i^\text Z\}$.
A classical analogue for quantum annealing is simulated annealing~(SA), which is a numerical global optimisation technique with ``temperature" guiding the simulated system towards a minimum energy state~\cite{KGV83}. 

Quantum annealing relies on the adiabatic evolution of a time-dependent Hamiltonian~\cite{FGGS00}
\begin{equation}
\label{eq:Hanneal}
H_{\rm QA}(t/t_\text f)
= - A({t/t_\text{f}})H_\text I + B({t/t_\text{f}}) H_\text P,
\end{equation}
for a duration~$t_\text{f}$, which is called the annealing time.
Ideally, the magnitude of $t_\text{f}$ is determined from the difference between ground and first excited energy levels of $H_{\rm QA}(t/t_\text f)$~\cite{FGGS00}.
Here~$A({t/t_\text{f}})$ and~$B({t/t_\text{f}})$ are smooth and monotonic functions defining a preset annealing schedule, and the initial Hamiltonian~$H_\text I$ is a trivial Hamiltonian satisfying $[H_\text I,H_\text P] \ne 0$.
At the beginning of an ideal quantum annealing process, the system starts in the ground state of $H_\text I$.
For a transverse field Hamiltonian~$H_\text I = \sum_{i} X_i$, its ground state is a uncoupled state, with each spin being in an equal superposition of $\sigma_i^\text Z=-1$ and $\sigma_i^\text Z=+1$.
During annealing, the system Hamiltonian~$H_{\rm QA}(t/t_\text f)$ slowly changes from $H_\text I$ to $H_\text P$ by decreasing $A(t/t_\text{f})$ smoothly from a maximum value to zero and increasing $B(t/t_\text{f})$ smoothly from zero to a maximum value.
At the end of the anneal, the system is ideally in the ground state of $H_\text P$, which encodes the solution of the given discrete optimisation problem.

The Ising minimisation problem is equivalent to the computational problem of QUBO, under the linear transformation $s_i\mapsto 2a_i-1$~\cite{BCMR10}.
A QUBO problem is finding the assignment of $\bm{a}$ that minimises the objective function
\begin{equation}
\label{eq:quboproblem}
    E(\bm a)=\bm{a}^\T \widetilde{\bm{Q}}\bm{a},\quad a_i\in\{0,1\},
\end{equation}
where $\bm{a}$ is a column vector of the binary variables~$a_i$ and the QUBO matrix $\widetilde{\bm{Q}}$ is a real-symmetric matrix. 
The diagonal and off-diagonal elements of $\widetilde{\bm{Q}}$ can be expressed as functions of~$h_i$ and~$J_{ij}$ up to a constant. 
Additionally, the QUBO problem~\eqref{eq:quboproblem}, or equivalently the Ising minimisation problem~\eqref{eq:ising_hamiltonian}, can be represented as an undirected graph~$G=\{V,E\}$.
The set of nodes~$V$ corresponds to the spin-$1/2$ particles with $\{h_i\}$ and $\{J_{ij}\}$ corresponding to the weights of nodes and edges~$E$, respectively.

The D-Wave System Inc.~offers a 5000-spin implementation of a practical quantum annealing device, commonly known as a quantum annealer or quantum processing unit~(QPU). 
The spins in this annealer are superconducting flux qubits, operating at a temperature of 15~mK, which are arranged in a Pegasus topology~\cite{DSC19}.
The native connectivity of the annealer chip \texttt{Advantage\_system}~1.1, which we use, has 5640 qubits (nodes) and 40484 couplers (edges), but a working chip typically has a fewer number of qubits and couplers due to technical imperfections.
Although this restricted connectivity only allows complete graphs of size less than 180 to be solved directly on the annealer~\cite{PHD19}, using the quantum-classical hybrid annealing solver~\texttt{hybrid\_binary\_quadratic\_model\_version2}~\cite{leaphybrid} we could solve up to a $10^6$-variable optimisation problem. 

D-Wave provides cloud-based access to its annealers using the quantum cloud service~\texttt{Leap}.
To make an optimisation problem compatible for a D-Wave solver, it needs to be first converted into an Ising minimisation problem~\eqref{eq:ising_hamiltonian} or a QUBO problem~\eqref{eq:quboproblem}.
Using predefined functions~\cite{CMR14}, the solver then embeds the problem into the Pegasus graph structure of the D-Wave annealer.
The Hybrid Solvers employ state-of-the-art classical algorithms, aided with automatic intelligent access of the quantum annealer, to deliver the best solution for the optimisation problem.
These solvers do not require precise manual controls of the annealers, making them suitable for various machine learning applications.
However, we can choose the value of the parameter~$\texttt{time\_limit}$, which denotes the maximum allowed problem runtime~\cite{timelimit}.

D-Wave is utilised for a regression task on lattice quantum chromodynamics simulation data, where the D-Wave annealer performed comparably to the best classical regression algorithm~\cite{NKY20}.
D-Wave is used to train a linear regression model about thrice faster than the classical approach with similar values for regression error metric, when applied on a synthetic dataset~\cite{DP20}.
Although no claims regarding a computational speedup over a classical soft-margin SVM is made, empirical evidence shows a better or comparable 
performance of quantum-annealing-based SVM in terms of classification accuracy, area under Receiver-Operating-Characteristic curve and area under Precision-Recall curve, for a dataset of size~$\approx1600$ in a binary classification problem of computational biology~\cite{WWDM20}.

\section{Approach}
\label{sec:approach}
In this section, we describe our approach for solving the regression problem of detecting facial landmarks using quantum-assisted $\varepsilon$-SVR models.
We begin by explaining our method to construct a quantum-assisted $\varepsilon$-SVR model and train it using a commercial quantum annealer.
Then we describe our FLD algorithm, which involves converting the multi-output regression problem into several single-output regression problems and solving each of them using SVR.
Finally, we discuss procedures to assess and compare classical and hybrid models for FLD.

\subsection{Quantum-assisted SVR}
\label{subsec:annealingSVR}
Here we explain our method for designing a quantum-assisted SVR, which involves employing quantum annealing for solving the dual formulation of the single-variable optimisation problem~\eqref{eq:minimize}.
First we show how to convert this optimisation problem into a QUBO problem~\eqref{eq:quboproblem}. 
Then we describe implementations of two $\varepsilon$-SVR models, constructed by solving the QUBO problem using a classical or a quantum-classical hybrid algorithm.
Finally, we briefly discuss the application of the quantum-assisted SVR formulation to the FLD problem, along with software packages utilised in this work.

Converting the optimisation problem in~\cref{eq:minimize} into a QUBO problem~\eqref{eq:quboproblem} is a two-step procedure: first we convert the constrained optimisation into an unconstrained optimisation on real values, and then we convert the real-valued unconstrained optimisation into a binary form.
To design an unconstrained optimisation problem, we construct a new objective function~$\mathcal{L}(\bm{\alpha})$ by adding the first constraint in~\cref{eq:minimize-2} as a square-penalty term to the objective function~\eqref{eq:minimize-1} using a Lagrange multiplier~$\lambda \in \Reals^+$~\cite{GKD18}. 
The solution of the unconstrained optimisation problem is a vector~$\bm{\alpha} \in\Reals^{2M}$.
For each element of~$\bm{\alpha}$, we define a binary encoding~\cite{WWDM20}, with total number of bits~$B$ and with~$B_\text{f}$ bits representing the fractional part, as
\begin{equation}
\label{eq:encoding}
\alpha_m \approx \frac{1}{2^{B_\text{f}}}
\sum_{i=0}^{B-1} 2^i a_{Bm+i}
\quad\forall m\in[2M],
\end{equation}
up to the encoding precision~$2^{-B_\text{f}-1}$.
By this encoding, we obtain an objective function~$E(\bm a)$, which is in the QUBO form~\eqref{eq:quboproblem}.
Additionally, with the choice for the regularisation hyperparameter 
\begin{align}
\label{eq:gamma}
    \gamma
    \approx \frac{1}{2^{B_\text{f}}}\sum_{i=0}^{B-1}2^i = \frac{1}{2^{B_\text{f}}}\left(2^{B}-1\right),
\end{align}
which, up to the encoding precision, is the maximum value possible for each~$\alpha_m$, the second constraint in Eq.~(\ref{eq:minimize-2}) is also satisfied.

We employ two different algorithms to solve the constructed QUBO problem, i.e., to minimise the objective function~$E(\bm{a})$~\eqref{eq:quboproblem}, and compare their performances.
One of the algorithms is purely classical and the other is a hybrid quantum-classical algorithm.
For a small training dataset of $M=100$ images and an encoding supporting only $B=5$ bits, the QUBO problem has 1000 variables with all non-zero quadratic terms, which cannot be solved using any pure quantum algorithm on current hardware.
We choose SA for classical optimisation and D-Wave's Hybrid Solver for quantum-classical hybrid optimisation, as SA and Hybrid are the typical choices in this field~\cite{HJK+20, PSH21}.
For SA, we use the implementation in D-Wave's package \texttt{dwave-neal}~\cite{DWaveNeal}, and use \texttt{LeapHybridSampler}~\cite{leaphybrid} for hybrid optimisation.
The output of each algorithm is a binary vector~$\bm{a}$, which after decoding by~\cref{eq:encoding} yields the solution~$\bm\alpha$ of the unconstrained optimisation problem in~\cref{eq:minimize}.
Having~$\bm \alpha$, we compute the offset~$b$~\eqref{eq:prediction_function_nl} by the method described in~\cref{appendix:offset}.
The vector~$\bm \alpha$ together with the offset~$b$ are then used to construct an $\varepsilon$-SVR model~\eqref{eq:prediction_function_nl}, 
where~$\varepsilon$ is the error tolerance in~\cref{eq:constrains}.
Henceforth, we refer to the $\varepsilon$-SVR models constructed by SA and Hybrid (quantum-assisted) solver as ``SA-SVR" and ``QA-SVR", respectively.

We apply this quantum-assisted $\varepsilon$-SVR formulation to the computational problem in FLD.
In the following subsections, we describe our approach for using QA-SVR and SA-SVR, and compare their performances for the FLD task~\eqref{eq:shape}.
We also compare these models against the standard $\varepsilon$-SVR model, which we refer to as ``SKL-SVR'', constructed using Python's \texttt{scikit-learn}~\cite{sklearnSVR}. 

\subsection{Algorithm for FLD}
\label{subsec:decomposition}
In this subsection, we elaborate on our algorithm for constructing and assessing SL models to solve the FLD task~\eqref{eq:shape}.
First we describe the raw datasets of facial images used in training and testing the models.
For images with~$L$ landmarks, we then split the FLD task~\eqref{eq:shape} into $2L$ sub-tasks, which corresponds to detecting $x$ and~$y$ coordinates separately, and construct an $\varepsilon$-SVR model~\eqref{eq:prediction_function_nl} for each sub-task.
Next we obtain a FLD model using models for the~$x$ and~$y$ coordinates of all landmarks.
Finally, we elaborate on our choice of performance metrics used for model evaluation.

\begin{figure}
    \centering
    \includegraphics[width=.65\linewidth]{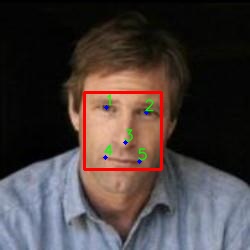}
    \caption{An example image from the LFW database~\cite{HRBL07}. Using data from Ref.~\cite{SWT13web}, we show the five landmarks with manually determined positions, labels 1--5, and the face box as a red rectangle. 
    }
    \label{fig:test_img}
\end{figure}

We use the datasets compiled from the publicly available databases LFW, LFPW and BioID for training and testing our SVR models~\cite{SWT13,SWT13web}.
The raw dataset~$\mathcal{D}^\text{raw}$~\eqref{eq:rawdata}, with~$N=125$ images and~$L=5$ landmarks, is a randomly selected subset of the LFW image database.
Each face shape is represented by a list of $(x,y)$ coordinates\footnote{Here the coordinate system is an inverted Cartesian system with the origin at the top-left corner of the image. This inverted coordinate system is the same as in \texttt{OpenCV} where images are represented as matrices.} of five facial landmarks:
left-eye centre~(1), right-eye centre~(2), nose tip~(3), left mouth corner~(4) and right mouth corner~(5); see~\cref{fig:test_img}.
In addition to the face shape, $\mathcal{D}^\text{raw}$ includes a representation of the face box,
shown in~\cref{fig:test_img},
which demarcates the extent of the face.
A face box is defined by a list of four numbers: first and third numbers are the coordinates of the top-left corner of the box, and the second and fourth numbers are the coordinates of the bottom-right corner.
Although these coordinates can vary depending on the face detection algorithm and can have biases based on age, gender, race, etc., in this paper we just use the pre-determined data~\cite{SWT13web}.

We design a FLD algorithm that solves the FLD task~\eqref{eq:shape} by splitting into $2L$ independent sub-tasks labelled by~$\ell\in[2L]$.
The raw dataset
\begin{equation}
\label{eq:rawdata_l}
  \mathcal{D}^{\text{raw}}_\ell=\{(\bm{I}^\text{raw}_i,s^{\text{raw}}_{i,\ell}) \mid i\in [N]\} \subset \Integers^{m\times n \times 3} \times \Reals,
\end{equation}
for sub-task~$\ell$, is a subset of $\mathcal{D}^{\text{raw}}$~\eqref{eq:rawdata}.
Given the above dataset, the sub-task is then to devise a model
\begin{equation}
\label{eq:shape_l}
\texttt{shape}_\ell: \Integers^{m\times n \times 3} \to \Reals : \bm{I}^\text{raw} \mapsto s^{\text{raw}}_\ell
\end{equation}
that accurately predicts $s^{\text{raw}}_\ell$ for an unmarked raw image~$\bm{I}^\text{raw}$.
By concatenating outputs of these $2L$ models~\eqref{eq:shape_l}, we obtain the face shape~$\bm{s}^\text{raw}$~\eqref{eq:shape_list} of~$\bm{I}^\text{raw}$.
Thus solving the $2L$ sub-tasks solves the FLD task of predicting the face shape.

Each sub-task~$\ell$~\eqref{eq:shape_l} involves a nonlinear single-output regression problem, which we solve by employing machine learning.
Using a dataset~$\mathcal{D}^{(\ell)}$, generated from $\mathcal{D}_\ell^\text{raw}$~\eqref{eq:rawdata_l} by preprocessing operations, the SL agent learns a regression model~$\widehat{\texttt{detect}}_\ell$, which predicts the scaled value of one coordinate of one landmark.
Due to the nonlinearity property and the high-generalisation capability of SVR~\cite{VMBP10}, we represent each $\widehat{\texttt{detect}}_\ell$ by an $\varepsilon$-SVR model~\eqref{eq:prediction_function_nl} with Gaussian kernel~\eqref{eq:KGaussian}.
This kernel function is typically used for non-linear regressions, and is also the default for the \texttt{kernel} parameter in \texttt{scikit-learn}.
Moreover, we compare classical and quantum-assisted algorithms by constructing three different $\varepsilon$-SVR models, namely QA-SVR, SA-SVR and SKL-SVR, for each $\widehat{\texttt{detect}}_\ell$.
The steps involved in the model construction are elaborated in \cref{appendix:workflow}.

We now construct a FLD model by combining the detection models for~$x$ and~$y$ components of each landmark.
The FLD model for the $k$th landmark is
\begin{equation}
\label{eq:detect_landmark}
   \widehat{\texttt{landmark}}_k = \left(\widehat{\texttt{shape}}_{2k},\, \widehat{\texttt{shape}}_{2k+1}\right),
\end{equation}
where $\widehat{\texttt{shape}}_{2k}$ and $\widehat{\texttt{shape}}_{2k+1}$ approximately predict the $x$ and $y$ components of landmark~$k$, respectively.
Similar to~\cref{eq:shape_compose}, each $\widehat{\texttt{shape}}_{2k}$ is constructed using the corresponding  $\widehat{\texttt{detect}}_{2k}$.
Consequently, a FLD model, which is a collection $\left(\widehat{\texttt{landmark}}_k, \forall k\in[L]\right)$, has three different formulations: QA-\texttt{landmark} is the hybrid model and SA-\texttt{landmark} and SKL-\texttt{landmark} are the classical models.

To evaluate the efficacy of a FLD model, we use the mean and variance of normalised detection error and the failure rate, calculated for all $L$ landmarks.
The output of each $\widehat{\texttt{landmark}}_k$ is a tuple~($\tilde x_k,\tilde y_k$), which are the predicted values for the~$x$ and~$y$ coordinates of the landmark relative to the top-left corner of the image.
Using these predicted coordinates, the true coordinates $(x_k,y_k)$ and the width of the face bounding box as $d_k$, we calculate MNDE$_k$~\eqref{eq:MNDE}, variance of $e_k$~\eqref{eq:DE} and FR$_k$~\eqref{eq:FR}, for~$e_\text{th}=0.1$.
Moreover, we benchmark the quantum-assisted model against the two classical models in two ways, namely, a k-fold cross validation on the 125 LFW images and testing on valid subsets of LFPW and BioID, as detailed in the next subsection.

\subsection{FLD model evaluation}
\label{subsec:evaluation}
Now we elaborate our procedure to evaluate and compare the classical and hybrid models for FLD.
Using the k-fold cross-validation technique, we first assess models trained and tested on the same database, i.e., LFW.
Additionally, we test our trained FLD models on two different databases, namely LFPW and BioID.
The hybrid model QA-\texttt{landmark} is compared with the classical models SA-\texttt{landmark} and SKL-\texttt{landmark} based on their performances over different test sets and time required for training the corresponding $\varepsilon$-SVR models.

We perform a 5-fold cross-validation assessment on the raw dataset~$\mathcal{D}^\text{raw}$ comprising 125 LFW images.
To this end, we use \texttt{scikit-learn}'s function~\texttt{KFold} to split this dataset equally into five consecutive sets or folds, where one fold serves as $\mathcal{D}^\text{raw}_\text{test}$ and the remaining four folds together serves as $\mathcal{D}^\text{raw}_\text{model}$.
We denote a pair ($\mathcal{D}^\text{raw}_\text{model}$, $\mathcal{D}^\text{raw}_\text{test}$) as one instance of our FLD problem.
For each of the five instances generated from $\mathcal{D}^\text{raw}$, we construct a FLD model using $\mathcal{D}^\text{raw}_\text{model}$ and evaluate its MNDEs and FRs on $\mathcal{D}^\text{raw}_\text{test}$.
As the calibration step is computationally expensive, we perform this only once and re-use the optimal hyperparameters for training in the other four instances.
The classical and hybrid models are compared based on their performances for each instance and the average performances of all five instances.

The (wall-clock) time for training an $\varepsilon$-SVR model can be used to compare classical and hybrid solutions for FLD.
As training involves solving an optimisation problem, the time required for training is proportional to the optimisation time, with annealing-based algorithms having an overhead for QUBO formulation.
We calculate and report the wall-clock times required for training SKL-SVR, SA-SVR and QA-SVR models, with both including and excluding the QUBO overhead.
Additionally, the QPU access time and Hybrid Solver's runtime obtained from D-Wave's cloud services are also presented.

To evaluate how our trained FLD models generalise, we test these models on valid subsets of two benchmarking databases for FLD, namely LFPW and BioID. 
These two are the commonly used databases for unconstrained images, with BioID having lesser variations in face pose, illumination and expression as compared to LFPW.
As our test datasets,  we choose subsets of sizes 164 and 1341 from the LFPW and BioID databases, respectively.

\section{Results}
\label{sec:results}
In this section, we present our results on the performance of our quantum-assisted algorithm for FLD and compare our algorithm with classical algorithms for FLD.
First we formulate the optimisation problem involved in training an $\varepsilon$-SVR model as a QUBO problem.
Then we compare the performances of the three FLD models developed here, namely SKL-\texttt{landmark}, SA-\texttt{landmark} and QA-\texttt{landmark}, using a subset of the LFW database.
Finally, using the LFPW and BioID databases, we compare these three FLD models. 

\subsection{QUBO formulation of SVR}
\label{subsec:QUBOformulation}
Here we construct a QUBO formulation for the constrained optimisation problem~\eqref{eq:minimize} used in training an $\varepsilon$-SVR model.
We begin by deriving a real-valued unconstrained optimisation problem corresponding to the constrained optimisation problem.
Then we establish an expression for the QUBO matrix, which is used to derive the QUBO objective function~$E(\bm a)$~\eqref{eq:quboproblem}.

For the real-valued unconstrained optimisation problem, we derive the objective function~$\mathcal{L}(\bm{\alpha})$ by adding the first constraint in~\cref{eq:minimize-2} as a square-penalty term to the objective function~\eqref{eq:minimize-1} using the Lagrange multiplier~$\lambda$.
We obtain the expression
\begin{align}
\label{eq:minimize_unconstrained1}
\mathcal{L}(\bm{\alpha}) := \frac12\bm{\alpha}^\T\bm{Q}\bm{\alpha} +\bm\alpha\cdot\bm{c} + \lambda\left(\sum_{m=0}^{M-1}\alpha_m-\sum_{m=M}^{2M-1}\alpha_m\right)^2,
\end{align}
for this real-valued objective function.
The solution of this unconstrained optimisation problem is the vector~$\bm{\alpha} \in\Reals^{2M}$ that minimises $\mathcal{L}(\bm\alpha)$ and satisfies the bounds imposed by the second constraint in Eq.~\eqref{eq:minimize-2}.

We now express the objective function~\eqref{eq:minimize_unconstrained1} of the unconstrained optimisation problem in a QUBO form~\eqref{eq:quboproblem}.
By employing the binary encoding~\eqref{eq:encoding}, we derive the $2MB\times2MB$ symmetric QUBO matrix~$\tilde{\bm Q}$, with elements
\begin{align}
\label{eq:QUBOmatrix}
    \widetilde{Q}_{Bn+i,Bm+j}
    = &
    \frac12\frac{2^{i+j}}{2^{2B_\text{f}}} Q_{nm}
    + \frac{2^{i}}{2^{B_\text{f}}}
    \updelta_{nm}\updelta_{ij} c_n
    +\lambda\frac{2^{i+j}}{2^{2B_\text{f}}}\nonumber\\
    &-2 \lambda\frac{2^{i+j}}{2^{2B_\text{f}}} \bar{\Theta}(m-M)\Theta(n-M)\nonumber\\
    &-2 \lambda\frac{2^{i+j}}{2^{2B_\text{f}}}
    \bar{\Theta}(n-M)\Theta(m-M),
\end{align}
where 
\begin{equation}
\label{eq:Heaviside}
    \Theta(n):=
    \begin{cases}
    0 & \text{if $n<0$},\\
    1 & \text{if $n\ge 0$}, 
    \end{cases}
\end{equation}
is the Heaviside step function and
\begin{equation}
\label{eq:flipHeaviside}
    \bar{\Theta}(n) := 1- \Theta (n) \quad \forall n \in \Integers.
\end{equation}

The QUBO matrix is then used to express the optimisation problem in the QUBO form as
\begin{equation}
\label{eq:qubo}
    E(\bm a)= \sum_{n,m=0}^{2M-1}\sum_{i,j=0}^{B-1}
    a_{Bn+i}\widetilde{Q}_{Bn+i,Bm+j}a_{Bm+j},
\end{equation}
see~\cref{appendix:qubo} for a detailed derivation of this QUBO problem.
Minimising~$E(\bm{a})$~\eqref{eq:qubo} produces a $(2MB)$-bit string~$\bm{a}$, which upon decoding consistent with~\cref{eq:encoding}, yields an approximate solution of the unconstrained optimisation problem~\eqref{eq:minimize}, up to the encoding precision.

\begin{figure}
    \centering
\includegraphics[width=.95\linewidth]{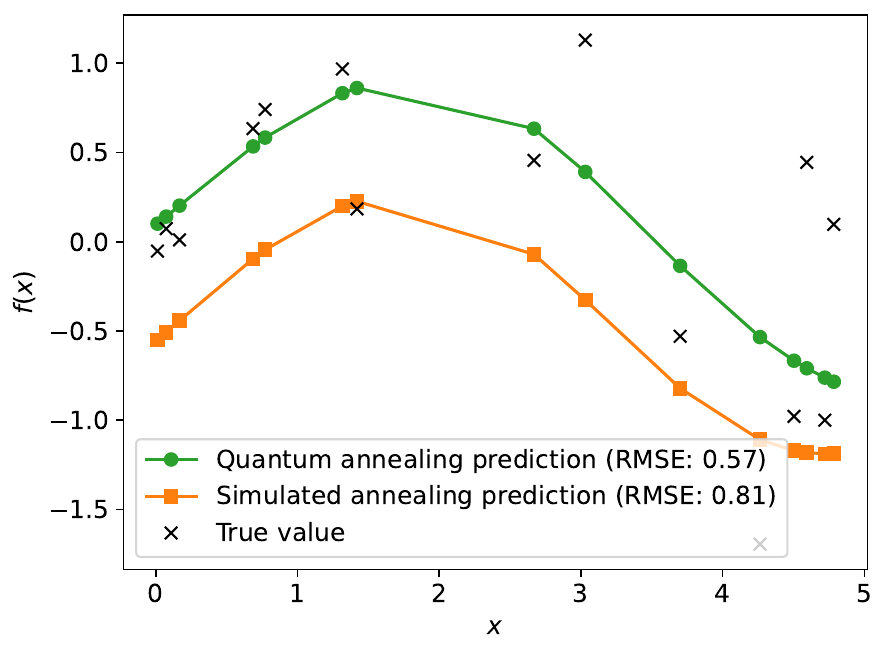}
    \caption{
    Performance comparison between simulated annealing and quantum annealing for a non-linear regression problem. The true values are generated by adding random noise to a sinusoidal function.
    We fix $\varepsilon=0.1$, $B=4$, $B_\text{f}=0$, $\lambda=1$ and $\eta=0.125$.
    Quantum annealing is run on D-Wave's \texttt{Advantage\_system4.1} with $0.5~\mu$s annealing time and 1000 repetitions, whereas simulated annealing uses 1000 sweeps and 1000 repetitions~(\cref{appendix:workflow}).
    RMSE is the root-mean-squared error between true and predicted values.
    }
    \label{fig:toy_result}
\end{figure}

Before applying the above QUBO formulation for the FLD problem, we test and compare simulated and quantum annealing solvers for a toy regression problem.
We use a simple curve-fitting example with 15 datapoints. By assuming an encoding using four binary variables for each real variable, we create a QUBO problem with 120 variables, which then fits the current 5000-qubit quantum annealer.
In Fig.~\ref{fig:toy_result}, we observe that the prediction error of quantum annealing is slightly lower than that of simulated annealing.
The predicted $f(x)$ from each solver is calculated as the average over 20 predictions from 20 different SVR models to account for the probabilistic nature of annealing.

\subsection{Quantum-assisted facial-landmark detection}
\label{subsec:fdl_SVR}
We compare the three implementations of our FLD model~\eqref{eq:detect_landmark}, namely, SKL-\texttt{landmark}, SA-\texttt{landmark} and QA-\texttt{landmark}, based on the resources required for training these models.
Each of these three FLD models is calibrated and trained using subsets of the LFW database~\cite{SWT13, SWT13web}.
Additionally, we show the true and predicted positions of five landmarks on an example LFW image.

We calculate the total wall-clock time and actual runtime for training an $\varepsilon$-SVR model $\widehat{\texttt{detect}}_\ell$ that constructs each component of the FLD model.
The annealing-based algorithms require an additional time for converting the dual optimisation problem for $\varepsilon$-SVR into a QUBO form~\eqref{eq:qubo}; we estimate this required time to be $\sim$5min for a problem with $M=100$.
Solving this QUBO problem using SA takes $\sim$25min of wall-clock time, as compared to $\sim$20s using D-Wave's Hybrid Solver.
More specifically, the actual runtime on this Hybrid Solver is $\sim$4s, with a QPU access time of $\sim$42ms.
On the other hand, it takes $\sim$8ms to train the SKL-SVR model.
These runtimes can vary depending on the classical and quantum hardware being used. The above runtime estimates are based on our available resources, namely, a 2.3 GHz Intel Core i5 processor and the D-Wave Solver \texttt{hybrid\_binary\_quadratic\_model\_version2} for classical and hybrid computations, respectively.

\begin{figure}
    \centering
    \includegraphics[width=.5\linewidth]{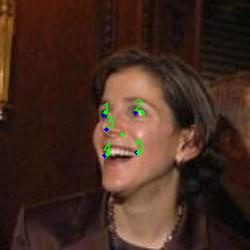}
    \caption[Compare sklearn, SA and Hybrid]{
    Predictions from QA-\texttt{landmark} model: actual landmark positions~\cite{SWT13web} (in blue) and predicted landmark positions (in green) for a LFW test image~\cite{HRBL07}.
    }
    \label{fig:LFW_results}
\end{figure}

\begin{figure}
    \centering
    \includegraphics[width=.85\linewidth]{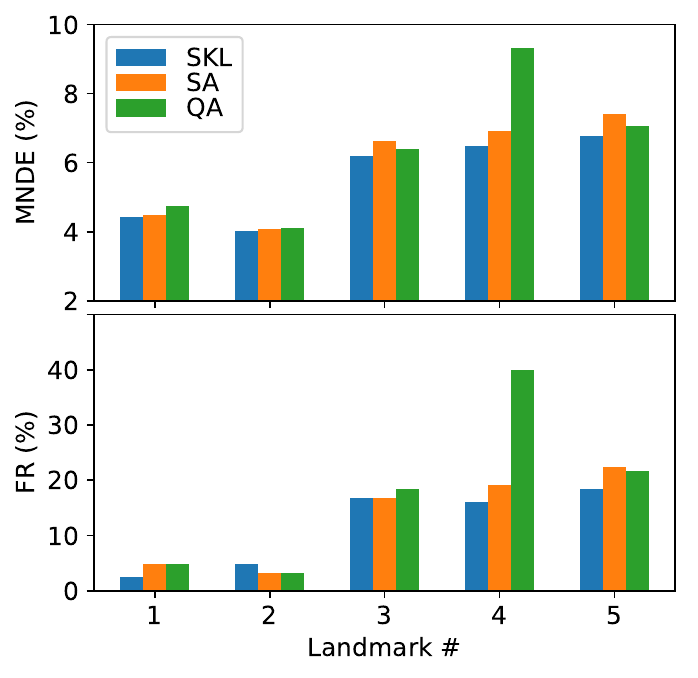}\\
    (a)\\
\includegraphics[width=.85\linewidth]{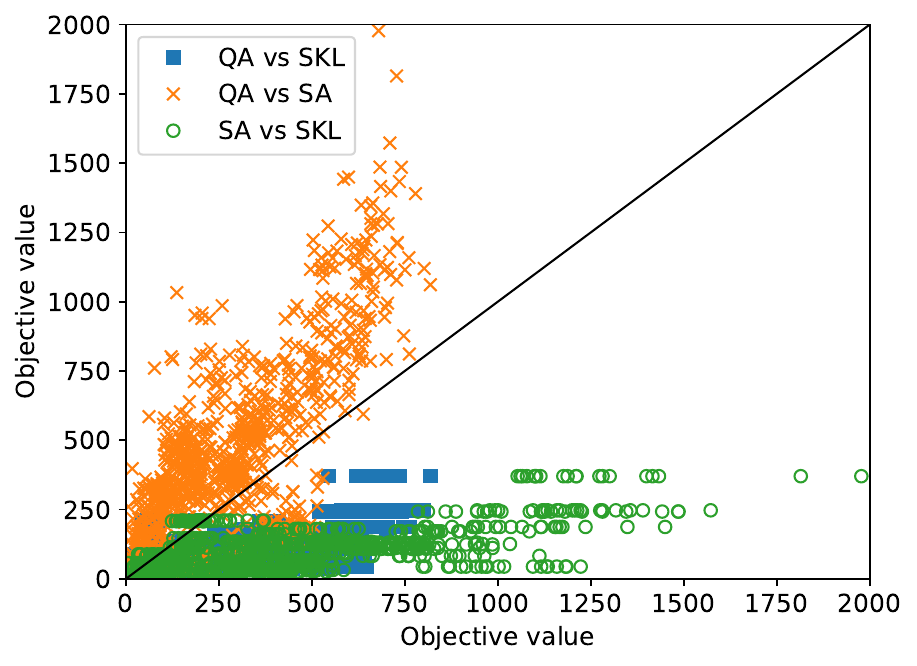}\\
    (b)
    \caption{
    Performance comparison between classical and quantum-classical hybrid optimisation techniques. 
    (a) For each model, namely SKL-\texttt{landmark}, SA-\texttt{landmark} and QA-\texttt{landmark}, and each landmark, we show as barplots the average (in \%) of five MNDEs and the average (in \%) of five FRs, which are obtained from the 5-fold cross validation on 125 LFW images.
    (b) For each pair of optimisers ($X$ vs $Y$) we plot data points~($O_X,O_Y$), where $O_X$ and $O_Y$ are the objective values obtained by the optimisers $X$ and $Y$, respectively. There are 1000 data points corresponding to 20 different SVR models for each landmark coordinate and each fold. The black diagonal line represents $O_X=O_Y$.
    }
    \label{fig:compare_1}
\end{figure}

By choosing the $5^\text{th}$ instance~(\cref{subsec:evaluation}) of our FLD problem,
we calibrate~(\cref{appendix:workflow}) each $\varepsilon$-SVR model to obtain optimal hyperparameter tuples; see \cref{tab:optimal} in \cref{app:results} for details.
In~\cref{fig:LFW_results}, we show detection results of the QA-\texttt{landmark} model, with optimal hyperparameters, for a LFW test image.
To assess the performance of our trained FLD model, we select an unconstrained LFW image, which has a non-frontal face with expression.
For the selected test image, the predicted positions of four landmarks~(\#1,~\#2,~\#4,~\#5) overlap with their corresponding true positions.
The predicted coordinates for landmark~\#3, i.e., tip of the nose, agree less to the actual coordinates by a normalised error~\eqref{eq:DE} of~19\%. 

\subsection{Benchmarking}
\label{subsec:performance_comparison}
We now evaluate and compare the efficacies of our three FLD models, namely SKL-\texttt{landmark}, SA-\texttt{landmark} and QA-\texttt{landmark}.
First we present results on our 5-fold cross validation of these three FLD models, followed by our results on model performances for unseen test datasets.
Finally, the performances of different trained FLD models over the aggregate of all five landmarks are presented.

In~\cref{fig:compare_1}(a), we plot the average of five MNDEs~\eqref{eq:MNDE} and the average of five FRs~\eqref{eq:FR} obtained from our 5-fold cross validation for each FLD model using 125 LFW images; see \cref{tab:k-fold} in \cref{app:results} for details.
QA-\texttt{landmark} delivers marginally lower MNDE than SA-\texttt{landmark} for landmarks~\#3 and~\#5, whereas MNDE for landmark~\#4 is $\sim$38\% higher than the two classical models.
These variations in MNDE are within the standard deviation $\sigma_e\approx0.034$ of each other~(\cref{tab:k-fold}).
All three FLD models yield FR$<5$\% for the first two landmarks, signifying that detection errors are below the threshold $e_\text{th}=0.1$ for over 95\% of our test images.
The QA-\texttt{landmark} model for landmark \#4 is significantly less successful than both classical detection models.

In order to further analyse performance of the classical and hybrid optimisers, we calculate the objective function values~\eqref{eq:optimisation_problem} using Lagrange multiplier solutions~\eqref{eq:normal_vector}.
From~\cref{fig:compare_1}(b), we observe that annealing-based methods usually fail to yield lower objective values as compared to the gradient-based method of \texttt{scikit-learn}.
The linear pattern in data points for `QA vs SKL' and `SA vs SKL' is due to the fact that for each landmark coordinate we compare 20 different objective values from the annealing-based method with only one from the gradient-based method. 
Furthermore, the quantum-assisted optimiser finds lower objective values than simulated annealing for most cases.

\begin{figure}
    \centering
    \includegraphics[width=.85\linewidth]{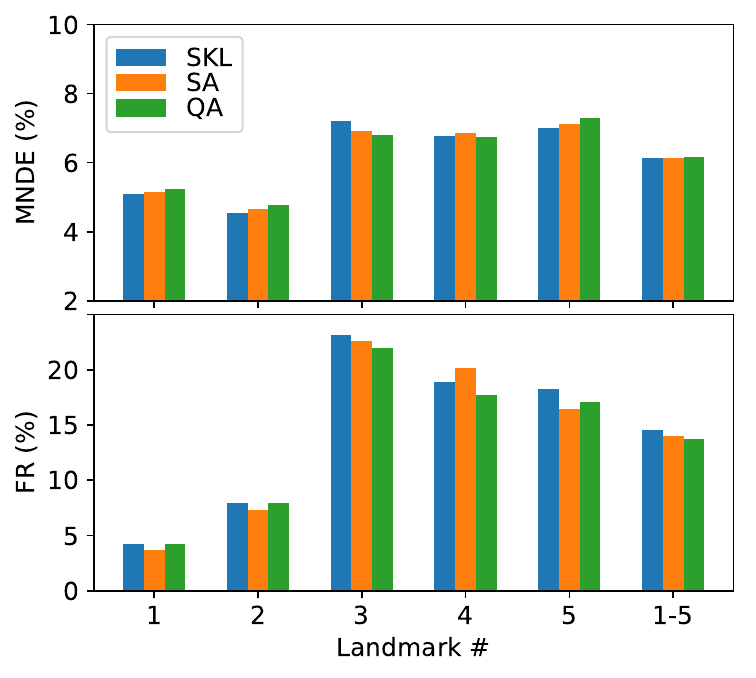}\\
    (a)\\
    \includegraphics[width=.85\linewidth]{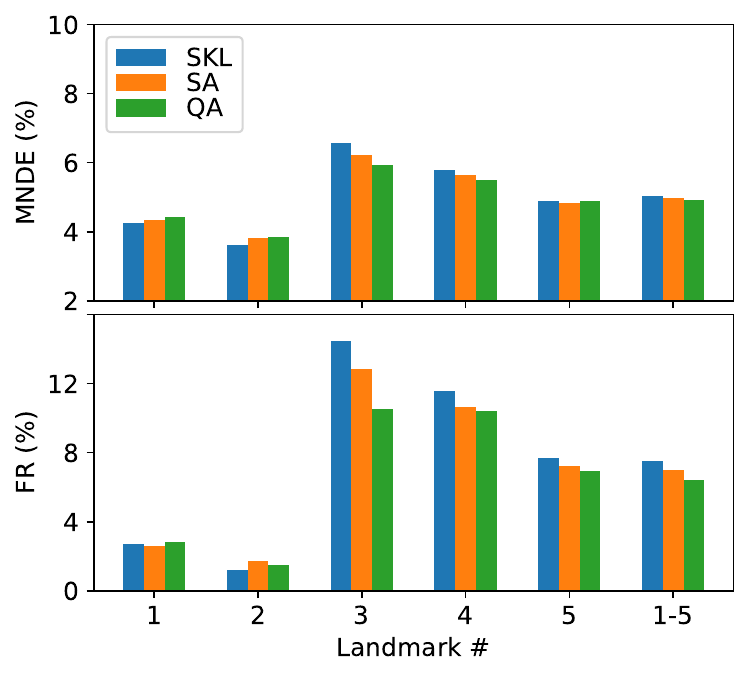}\\
    (b)
    \caption[Compare with state-of-the-art regression algorithms]{
    Performance comparison between SKL-\texttt{landmark}, SA-\texttt{landmark} and QA-\texttt{landmark}.
    The training dataset for (a) and (b) is the training set obtained in the $5^\text{th}$ fold of our 5-fold cross validation.
    For each landmark and the aggregate of all five landmarks, we plot the MNDE (in \%) and FR (in \%) for the test dataset comprising 164 LFPW images in (a) and for the test dataset comprising 1341 BioID images in~(b).
    }
    \label{fig:compare_2}
\end{figure}

We compare performances of the three FLD models, which are constructed using training datasets in the $5^\text{th}$ instance~(\cref{subsec:evaluation}), for the LFPW and BioID test datasets~\cite{SWT13, SWT13web}.
We choose the $5^\text{th}$ instance of our FLD problem because it was previously used for tuning the hyperparameters.
In Figs.~\ref{fig:compare_2} (a) and (b), we observe that the performances of all three FLD models are comparable for each landmark as well as on their aggregate.
Similar to the LFW results, MNDE and FR vales are lower for centers of both eyes as compared to the other three landmarks.
For BioID images, QA-\texttt{landmark} is slightly ($<20\%$) better than SA-\texttt{landmark} and SKL-\texttt{landmark} over the aggregate of all five landmarks,
but all MNDE values are within $\sigma_e$ of each other; see~\cref{tab:aggregate}.
We provide detailed results for LFPW and BioID datasets in Tables~\ref{tab:lfpw_results} and \ref{tab:bioid_results}, respectively.

We perform additional tests using the other four training datasets for the $1^\text{st}$ to $4^\text{th}$ instances and report performances of the aggregate case in~\cref{tab:aggregate}.
All trained FLD models perform better, with lower MNDEs and FRs for the aggregate of all five landmarks, on the BioID dataset as compared to the LFW and LFPW datasets.
In the rows corresponding to Fold \#5, where the training dataset is the same as in the $5^\text{th}$ instance, we notice that annealing-based algorithms consistently yield lower $\sigma_e$ than scikit-learn's technique. 
This advantage is absent for the other instances, whose hyperparameters are not optimised in our experiments.
\begin{table*}[ht!]
\begin{tabular}{|c|c||c|c|c|c|c|c|c|c|}
\hline 
 \multicolumn{2}{|c|}{Model} & \multicolumn{2}{|c|}{Testset 1 (LFW)}
 & \multicolumn{2}{|c|}{Testset 2 (LFPW)}
 & \multicolumn{2}{|c|}{Testset 3 (BioID)}\\
\hline 
 Fold \# & Type & \% MNDE ($\sigma_e$)& FR & \% MNDE ($\sigma_e$) & FR &
 \% MNDE ($\sigma_e$) & FR \\
\hline
\hline
 & SKL & 6.13 (0.0398) & 19.2 & 6.28 (.0387) & 16.71 & 5.08 (.0306) & 6.67\\
1 & SA & 6.34 (0.0417) & 16.8 & 7.38 (.0479) & 24.39 & 5.59 (.0330) & 10.54\\
 & QA & 6.8 (0.0424) & 20 & 6.69 (.0424) & 21.46 & 5.36 (.0334) & 8.72\\
\hline 
 & SKL & 5.59 (0.0416) & 14.4 & 6.54 (.0390) & 17.19 & 5.58 (.0333) & 10.72\\
2 & SA & 6.09 (0.0413) & 15.2 & 7.01 (.0414) & 22.44 & 5.79 (.0327) & 10.86\\
 & QA & 6.9 (0.0509) & 20.8 & 7.63 (.0459) & 27.19 & 6.62 (.0385) & 18.02\\
\hline
 & SKL & 5.08 (0.0295) & 4.8 & 6.51 (.0383) & 16.22 & 5.19 (.0317) & 8.01\\
3 & SA & 5.62 (0.0344) & 8.8 & 6.96 (.0412) & 20.24 & 5.48 (.0316) & 8.72\\
 & QA & 6.07 (0.0371) & 16.8 & 7.61 (.0451) & 26.22 & 6.47 (.0385) & 16.72\\
\hline
 & SKL & 5.88 (0.0330) & 11.2 & 6.10 (.0356) & 13.78 & 4.97 (.0333) & 6.35\\
4 & SA & 6.32 (0.0353) & 16 & 6.88 (.0392) & 19.51 & 5.76 (.0319) & 9.45\\
 & QA & 6.74 (.0393) & 22.4 & 7.11 (.0413) & 22.32 & 6.13 (.0378) & 14.49\\
\hline
 & SKL & 5.21 (0.0336) & 8.8 & 6.12 (.0373) & 14.51 & 5.02 (.0318)& 7.52\\
5 & SA & 5.14 (0.0328) & 9.6 & 6.14 (.0363) & 14.02 & 4.97 (0.0313) & 7\\
 & QA & 5.1 (0.0324)& 8 & 6.17 (.0365) & 13.78 & 4.92 (.0306)& 6.44\\
\hline
\end{tabular}
\caption{Overall results for all 15 instances (5 training sets with 3 test sets for each).
For each row, the training dataset is created using all data points in $\mathcal{D}^\text{raw}$ except the points in the corresponding Fold \#.
Each performance value is calculated for the aggregate of all 5 landmarks.
The values in the final row are used for plotting Figs.~\ref{fig:compare_2} (a) and (b).}
\label{tab:aggregate}
\end{table*}

\section{Discussion}
\label{sec:discussion}
In this section, we discuss the results of implementing our quantum-assisted algorithm for detecting facial landmarks on a practical quantum annealer.
We begin by analysing the derived unconstrained formulation for the optimisation problem involved in generating an $\varepsilon$-SVR model.
Then we elaborate on the quantum-classical hybrid implementation of our algorithm.
Finally, we analyse the performance of our FLD algorithm.

We construct a quantum-assisted $\varepsilon$-SVR model using D-Wave's quantum annealer.
For the purpose of generating this model, we cast the constrained optimisation problem~\eqref{eq:minimize} into a QUBO problem~\eqref{eq:quboproblem} by first deriving an unconstrained optimisation problem and then expressing this optimisation problem over binary variables.
The derived QUBO objective function~\eqref{eq:qubo} is equivalent to the original objective function~\eqref{eq:minimize-1} up to a judicious choice of the Lagrange multiplier~$\lambda$ and the encoding parameters~\eqref{eq:encoding}.
Due to imprecision of this real-to-binary encoding, our $\varepsilon$-SVR formulation is approximate.

We train the quantum-assisted $\varepsilon$-SVR model using D-Wave's Hybrid Solver, which uses a combination of classical algorithms and quantum annealing for optimising a QUBO problem.
Due to restrictions on the number and connectivity of qubits on a D-Wave QPU, the size of a fully connected graph that can be directly embedded onto the hardware is 180 for the 5000-qubit chip.
This limitation has led to the use of batch learning approaches for previous machine learning tasks on D-Wave annealers~\cite{LFRL18, WWDM20}.
In this work, we thus make use of the Hybrid Solver for SL, which has two benefits: solving QUBO problems with a million variables and bypassing the QPU's hyperparameter optimisation requirement.
Additionally, we observe that training $\varepsilon$-SVR models using Hybrid Solver is easier, faster and results in better performance than using a limited-size QPU.

Although both SVR and SVM are kernel-based techniques, they are fundamentally different in their model construction and applications.
In contrast to the previous work on quantum-annealing-based SVM~\cite{WWDM20}, the QUBO problem in our quantum-assisted SVR has a different objective function, with double the number of binary variables and one extra hyperparameter.
The bias term in the prediction function is also calculated differently in this work.
Additionally, we implement our SVR models using D-Wave's Hybrid Solver as opposed to the 2000-qubit quantum annealer used for training in Ref.~\cite{WWDM20}.
Thus, besides presenting a feasible use-case for quantum annealing, more generally quantum optimisation, our work assesses D-Wave's state-of-the-art solvers for kernel-based learning methods.

We employ the quantum-assisted $\varepsilon$-SVR model for efficient detection of facial landmarks, and assess its efficacy relative to both classical implementations, i.e., SA and \texttt{scikit-learn}, of our FLD algorithm.
These three FLD models are trained using a dataset of 100 LFW images.
On applying the quantum-assisted FLD model on an example LFW image, we observe that the predicted positions of four landmarks, namely eyes and mouth corners, are in good agreement with their corresponding true values, whereas the detection of the nose tip failed according to the standard failure threshold~\cite{CUS13}. 
Based on their average performances from a 5-fold cross-validation on 125 LFW images, the three FLD models are equivalent within statistical variations.
Although the hybrid technique and SA perform comparably, the hybrid optimisation is about $75\times$ faster than the classical optimisation.
Nevertheless, the \texttt{scikit-learn}'s SVR is much faster than the other two implementations, which employ global optimisation as compared to a gradient-based optimiser used in \texttt{scikit-learn}.
With further inspection, we observe that the hybrid technique frequently yields lower values for the SVR objective function as compared to those obtained by SA. This slight advantage can be attributed partly to quantum annealing.
On the other hand, \texttt{scikit-learn}'s gradient-based method tends to find even lower objective values than the annealing-based methods because with QUBO encoding of the real-valued optimisation problem, the search space gets degraded for the annealing-based methods.

We compare the three implementations of our FLD algorithm by training FLD models using the training dataset of the 5th fold, and subsequently testing them on the benchmarking datasets of LFPW and BioID~\cite{SWT13web}. 
Our choice of this fold is justified because we have previously used this particular fold for hyperparameter tuning.
Although we observe a slight advantage of the quantum-assisted models over all landmarks, we interpret this advantage as a statistical fluctuation and not a quantum advantage, as is claimed for similar works~\cite{LFRL18, WWDM20}.
The annealing-based implementations generate more accurate $\varepsilon$-SVR models with lower variances as compared to \texttt{scikit-learn}.
We contribute this advantage to the fact that the SVR model generated using SA or hybrid technique is an average of 20 feasible models, thus yielding a lower variance of MNDE for the test dataset. 
By choosing different starting points, we could generate multiple FLD models by a gradient-based approach as well, but this analysis is beyond the scope of our work.

\section{Conclusions}
\label{sec:conclusion}
We have adapted SVR, a popular tool in supervised learning, into a quantum-assisted formulation.
Our formulation employs quantum annealing for solving the optimisation problem, which is used to train the SVR model, with high accuracy.
We have constructed a quantum-assisted SVR model using D-Wave's Hybrid Solver and utilised this model for detecting five facial landmarks:
centres of both eyes, tip of the nose and corners of the mouth.
Furthermore, we tested efficacy by comparing landmark predictions of this model to predictions obtained from two classical models.

We have chosen the problem of FLD because it plays a key role in face recognition by assisting the conversion of unconstrained images to constrained images.
Recent FLD algorithms, which are based on neural networks and regression techniques, yield the best detection accuracies so far.
However, the success of these algorithms depends on the quality of available training datasets and the available computational resources.
As training an efficient and robust FLD model using a finite dataset of unconstrained images is still challenging for classical FLD algorithms, exploring quantum-assisted alternatives is thus worthwhile.

Quantum-assisted algorithms based on quantum annealing are shown to be empirically advantageous over classical algorithms for a variety of machine-learning problems.
Notable examples include the protein-binding problem in computational biology and the Higgs particle-classification problem in high-energy physics.
For these problems, quantum-assisted algorithms yield classifiers, trained using a small dataset, with superior accuracies compared to classical algorithms.
Quantum-algorithmic performance is deleteriously affected by practical limits, such as device noise, few qubits and restricted qubit connectivity.

We have proposed a quantum-assisted regression algorithm for the FLD task and tested this algorithm's performance using D-Wave's Hybrid Solver.
Our first result is a QUBO formulation for SVR.
Specifically, we derive a QUBO form for the constrained optimisation problem involved in training a SVR model. 
Our second result is a SVR-based FLD algorithm, which solves the multi-output regression task by splitting it into several single-output regression tasks and constructing a SVR model for each such single-output regression problem.
Upon implementing this algorithm on D-Wave's Hybrid Solver, we have observed comparable performance against classical implementations in terms of FLD accuracy.
Furthermore, we notice that annealing-based FLD algorithms yield solutions with lower variances than those obtained using gradient-based algorithms.

Our work is a proof-of-concept example for applying quantum-assisted SVR to a real-world supervised learning task.
Although we study the variance of our results to conclude a slight advantage of annealing-based methods over gradient-based methods, higher-order statistical fluctuations need to be analysed.
Some of the possible improvements to the implementations of our FLD algorithm include increasing number of image segments during feature extraction, optimising over annealing hyperparameters 
and exploring customised workflows for Hybrid Solver.
Moreover, future experiments on a larger quantum annealer with exclusive QPU access, instead of hybrid optimisation schemes,  
might have the potential to yield statistically significant quantum advantage.

During the long refereeing process of our work, related publication emerged that formulates a similar quantum SVR model to estimate chlorophyll concentration in water~\cite{PRM+22}.
In particular, both works apply quantum (and hybrid) annealing to solve the constrained optimisation problem involved in SVR training, but the constraint handling is done in a slightly different manner.
Additionally, there are some other notable differences in the ML workflow including the hyperparameter tuning step and benchmarking.

\section*{Data availability}
All data generated or analysed during this study are included in this article and its Appendix~\ref{app:results}.

\section*{Acknowledgements}
We acknowledge the traditional owners of the land on which this work was undertaken at the University of Calgary: the Treaty 7 First Nations. 
We appreciate Dr.\ J\"org Denzinger for his detailed and critical comments on our manuscript.

\section*{Declarations}
\textbf{Funding:} This work is supported by the Major Innovation Fund, Government of Alberta, Canada.

\noindent
\textbf{Conflict of interest:} The authors declare no competing interests.

\bibliography{paper}

\appendix
\section{SVR basics}
\label{appendix:svr_basics}
Here we discuss the primal formulation of a linear SVR. 
First we describe SVR as a tool for solving the supervised-learning problem of regression. Then we discuss the associated constrained optimisation problem required in training a SVR model.

\begin{figure}[hb!]
    \centering   \includegraphics[width=.45\linewidth]{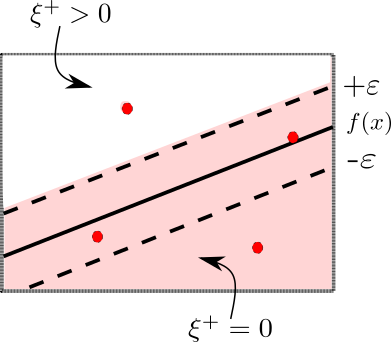}  \includegraphics[width=.45\linewidth]{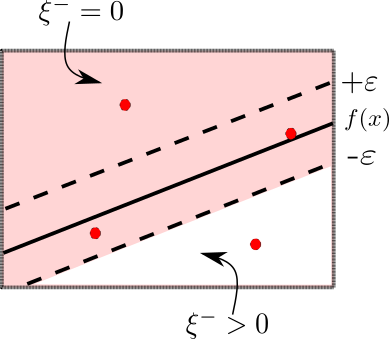}
    \caption[Slack variables for $\varepsilon$-SVR]{Two slack variables~$\xi^+$ and~$\xi^-$ are used in an $\varepsilon$-SVR formulation.
    Red circles are data points.
    (left)~$\xi^+=0$ if the corresponding training data point is below the upper bound and $\xi^+>0$ if it is above the upper bound,
    (right)~$\xi^-=0$ if the corresponding training data point is above the lower bound and $\xi^->0$ if it is below the lower bound.}
    \label{fig:scalc_variables}
\end{figure}

The linear $\varepsilon$-SVR is formally written as the convex-optimisation problem
\begin{equation}
\label{eq:optimisation_problem}
    \min_{\bm{w},b} \left\{\left.\frac{\bm{w}^2}2 \right\vert \abs{\bm{w}\cdot\bm{x}_i+b-y_i}\leq \varepsilon \quad \forall i\in[M]\right\}.
\end{equation}
This optimisation problem might not be feasible; i.e.
it is possible that no linear function~$f(\bm{x})$~\eqref{eq:linear_function} exists to satisfy the constraint in~\cref{eq:constrains} for all training data points~\eqref{eq:train_data}.
In order to cope with this infeasibility issue, analogous to the soft-margin concept in support vector machine classification~\cite{SS18}, slack variables are used.
Specifically, in the soft-margin $\varepsilon$-SVR, two slack variables $\bm{\xi}^+,\bm{\xi}^-\in (\Reals^+)^M$ are introduced for each training point
\footnote{If we use one slack variable then the constraints in the optimisation problem would be absolute value of some function and therefore the constraints would be non-differentiable.
In this case deriving the dual formulation or the KKT conditions will be cumbersome.};
see~\cref{fig:scalc_variables}.

Introducing these slack variables and a regularisation hyperparameter~$\gamma~\in~\Reals^+$ leads to the optimisation problem
\begin{equation}
\label{eq:primal}
    \min_{\stackrel{\bm{w},b}{\bm{\xi}^+, \bm{\xi}^-}}\left\{
    \frac{\bm{w}^2}2
    +\gamma \norm{\bm\xi^+ + \bm\xi^-}_1
    \right\},
\end{equation}
subject to the inequality constraints
\begin{equation}
\label{eq:primalconstraints}
-\varepsilon-\xi^+_i
\leq \bm{w}\cdot\bm{x}_i + b - y_i \leq \varepsilon + \xi^-_i
\quad
\forall i \in [M].
\end{equation}
The optimisation problem in~\cref{eq:primal} is known as the primal formulation of $\varepsilon$-SVR.
The regularisation hyperparameter~$\gamma$ in this equation determines trade-off between minimising the norm~$\norm{\bm w}$, i.e., the flatness of the function $f(\bm{x})$, and the amount by which deviations greater than the error~$\varepsilon$ are tolerated.

\section{Computing the offset in prediction function}
\label{appendix:offset}
In this appendix, we provide a comprehensive overview of the methods used in the \texttt{LIBSVM} library~\cite[p.~10]{CL11} for computing the offset~$b$ in the prediction function~\eqref{eq:prediction_function_nl}.
In this method, the Karush-Kuhn-Tucker~(KKT) conditions are employed to derive bounds for~$b$ and estimate its value.
We begin by discussing the KKT conditions and their implications for a linear prediction function~\eqref{eq:prediction_function}.
These conditions yield general bounds for $b$.
Finally, we provide two methods used for estimating $b$ in practice.

The KKT conditions are constraints required to obtain optimal solutions.
These conditions govern the relations between the dual variables $(\bm{\alpha}^+,\bm{\alpha}^-)$ and the constraints in the primal formulation~\eqref{eq:primalconstraints}.
Given a data point~$(\bm{x}_i, y_i)$ and a particular solution $f_\text{linear}(\bm{x}_i)=\bm {w}\cdot\bm{x}_i+b$, the KKT conditions are~\cite[Eqs.~(12,13)]{SS04}
\begin{align}
    \alpha^+_i\left(\varepsilon+\xi^+_i-y_i+\bm {w}\cdot\bm{x}_i+b\right) &= 0 
    \label{eq:KKT-1},\\
    \alpha^-_i\left(\varepsilon+\xi^-_i+y_i-\bm {w}\cdot\bm{x}_i-b\right) &= 0 \label{eq:KKT-2},\\
    \xi^+_i\left(\gamma-\alpha^+_i\right) &= 0 \label{eq:KKT-3},\\
    \xi^-_i\left(\gamma-\alpha^-_i\right) &= 0.
    \label{eq:KKT-4}
\end{align}

By~\cref{fig:scalc_variables}, we construct mathematical conditions for the data point to be inside or outside the $\varepsilon$-insensitive tube.
Inside the tube, the data point can be either above or below the line representing the prediction function, as expressed by the relations
\begin{equation}
\label{eq:inside_tube_up}
  y_i-f_\text{linear}(x_i) \le \varepsilon, \quad
  \xi^+_i = 0,
\end{equation}
and
\begin{equation}
\label{eq:inside_tube_down}
  -\varepsilon \le y_i-f_\text{linear}(x_i), \quad
  \xi^-_i = 0,
\end{equation}
respectively.
If the point resides outside the tube, it can either be above the upper margin, with
\begin{equation}
\label{eq:outside_tube_up}
 y_i-f_\text{linear}(x_i)-\varepsilon \ge 0,
\end{equation}
or below it, for 
\begin{equation}
\label{eq:outside_tube_down}
 y_i-f_\text{linear}(x_i)+\varepsilon \le 0.
\end{equation}

Using the set of KKT conditions, one can derive the upper and lower bounds for $b$.
For $\alpha^+_i<\gamma$, $\xi^+_i=0$ by Eq.~\eqref{eq:KKT-3}, and the corresponding datapoint is inside the tube~\eqref{eq:inside_tube_up}.
Additionally, $\alpha^+_i>0$ implies that the expression inside parenthesis in Eq.~\eqref{eq:KKT-1} needs to be zero and, consequently, the point is outside the tube~\eqref{eq:outside_tube_up}.
Similarly for $\alpha^-_i<\gamma$, we can infer that $\xi^-_i=0$~\eqref{eq:KKT-4} and the point is inside the tube~\eqref{eq:inside_tube_down}.
For the case when $\alpha^-_i>0$, Eq.~\eqref{eq:KKT-2} suggests that the point is outside the tube and below the lower margin~\eqref{eq:outside_tube_down}.
Based on these observations, we can concisely state the bounds for $b$ as~\cite{CL11}
\begin{align}
\label{eq:bound_b}
    &\max\{b_i^- \mid \alpha^+_i<\gamma \text{ or } b_i^+ \mid \alpha^-_i>0\} \leq b\leq \nonumber\\
    &\min\{b_i^-\mid \alpha^+_i>0 \text{ or } b_i^+ \mid \alpha^-_i <\gamma\}
    \quad \forall i \in [M],
\end{align}
with
\begin{equation}
b^{\pm}_i:=\pm\varepsilon+y_i
-\sum_{j=0}^{M-1}(\alpha_j^+-\alpha_j^-)
\,\bm{x}_j\cdot\bm{x}_i.
\end{equation}

We now describe the methods used in the \texttt{LIBSVM} library~\cite[p.~10]{CL11} for computing~$b$.
If there exists at least one $\alpha^+_i$ or $\alpha^-_i$ that lies in the interval~$(0,\gamma)$, the inequalities in~\cref{eq:bound_b} become equalities, and $b$ is estimated as the average
\begin{equation}
\label{eq:offset1}
    b= \frac{\displaystyle\sum_{i: \alpha^+_i \in(0,\gamma)} b^-_i +\sum_{i: \alpha^-_i \in(0,\gamma)} b^+_i}{\abs{\{i\mid \alpha^+_i \text{ or } \alpha^-_i \in (0, \gamma)\}}}.
\end{equation}
For the case where no~$\alpha^+_i$ or~$\alpha^-_i$ is in the interval~$(0,\gamma)$, the bounds in~\cref{eq:bound_b} simplifies to
\begin{align}
\label{eq:offset2}
    &\max\{b^-_i \mid \alpha^+_i=0 \text{ or } b^+_i \mid \alpha^-_i=\gamma\} \leq b\leq \nonumber\\
    &\min\{b^-_i\mid \alpha^+_i=\gamma \text{ or } b^+_i \mid \alpha^-_i =0\},
\end{align}
and~$b$ is estimated to be the midpoint of this range.
In the non-linear prediction function~\eqref{eq:prediction_function_nl}, we use $K(\bm{x}_j,\bm{x}_i)$ instead of $\bm{x}_j.\bm{x}_i$ in the calculations of $b^+_i$ and $b^-_i$.

\section{Preprocessing images}
\label{appendix:preprocess}
The normalisation operation proceeds by first converting the raw images into grayscale images and detecting the facial region within each grayscale image by a face-detection algorithm such as the Viola–Jones algorithm~\cite{VJ01}.
The detected face region is then cropped and converted into a common-size image.
We describe the normalisation process for each image by the map
\begin{equation}
\label{eq:normalise}
    \texttt{normalise}: \Integers^{m\times n \times 3} \to \Integers^{m_\text{r}\times n_\text{r}}:
    \bm{I}^\text{raw} \mapsto \bm{I}^\text{norm},
\end{equation}
where $m_\text{r} \cross n_\text{r}$ is the size of grayscale images after normalisation and $\bm{I}^\text{norm}$ denotes a normalised image.
By normalisation, the face shape $\bm{s}^\text{raw}$ of each raw image is scaled by the dimension $(m_\text{r}$, $n_\text{r})$ of the normalised image according to
\begin{equation}
\label{eq:scale}
    \texttt{scale}: \Reals^{2L} \to \Reals^{2L}: \bm{s}^\text{raw} \mapsto \bm{s},
\end{equation}
where~$\bm{s}$ denotes the face shape of a normalised image.
Thus, normalisation of marked images involves two functions, namely \texttt{normalise} and \texttt{scale}, acting on images and their corresponding landmarks, respectively.

Feature extraction is performed by a feature descriptor
\begin{equation}
\label{eq:extract}
    \texttt{extract:} \Integers^{m_\text{r}\times n_\text{r}} \to \Reals^{F_\text{norm}}: \bm{I}^\text{norm} \mapsto \bm{x}^\text{norm},
\end{equation}
that maps a normalised image~$\bm{I}^\text{norm}$ into a feature vector~$\bm{x}^\text{norm}$ of size $F_\text{norm}$, which describes the normalised image.
The common feature descriptors are the Haar-like feature descriptor~\cite{VMBP10} and the local binary patterns (LBP) feature descriptor~\cite{MVBP13}.
For more details, we also refer to their implementations in Python's \texttt{scikit-image} library in~\cite{hlf} for the Haar-like descriptors and in~\cite{lbp} for the LBP descriptors. 
To overcome the overfitting problem due to high dimensionality of feature vectors, $F_\text{norm}$ is further reduced by feature selection techniques such as Adaboost regression~\cite{VMBP10} and correlation-based feature selection~(CFS)~\cite{CWWS14}.
A feature selection is a map
\begin{equation}
\label{eq:select}
    \texttt{select:} \Reals^{F_\text{norm}} \to \Reals^{F}: \bm{x}^\text{norm} \mapsto \bm{x},
\end{equation}
which maps a high-dimensional feature vector $\bm{x}^\text{norm}$ of size $F_\text{norm}$ to a low-dimensional feature vector $\bm{x}$ of size $F$.
We define all the operations in the preprocessing step as a composite function
\begin{equation}
\label{eq:preprocess}
    \texttt{preprocess}=\texttt{select} \circ \texttt{extract} \circ \texttt{normalise},
\end{equation}
which maps a raw image~$\bm{I}^\text{raw}$ to a preprocessed feature vector~$\bm{x}$.

\section{Constructing a model}
\label{appendix:workflow}
\begin{figure}
    \centering
    \includegraphics[width=\linewidth]{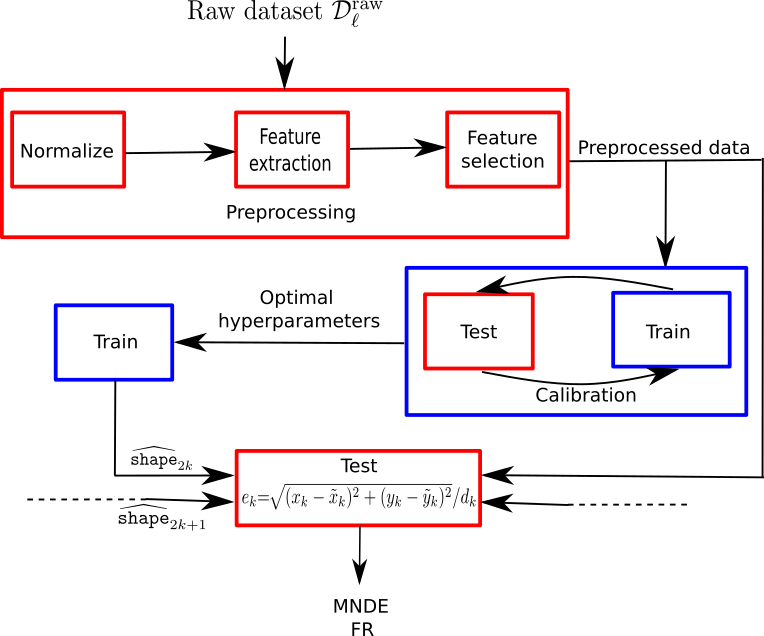}
    \caption{ML workflow for constructing and characterising a FLD model.
    The subscript~$\ell$ denotes one coordinate of a landmark~$k$.
    The red boxes represent purely classical operations, and each blue box denotes an operation that can be either classical, quantum or a hybrid of both.}
    \label{fig:MLworkflow}
\end{figure}

Here we explain our machine learning workflow~(\cref{fig:MLworkflow}) for constructing a model~$\widehat{\texttt{shape}}_\ell$, which approximates the ideal model~\eqref{eq:shape_l} for a sub-task~$\ell$.
We begin by describing the preprocessing operations applied on the raw dataset~$\mathcal{D}_\ell^\text{raw}$~\eqref{eq:rawdata_l}, which can be splitted as~\eqref{eq:ModelTest}
\begin{equation}
\label{eq:ModelTest_l_raw}
    \mathcal{D}_\ell^\text{raw}=\mathcal{D}_{\ell,\text{model}}^\text{raw}\sqcup \mathcal{D}_{\ell,\text{test}}^\text{raw}.
\end{equation}
Using the resultant processed dataset, we then construct an $\varepsilon$-SVR model, which approximately predicts one coordinate of one landmark for a normalised image.
In this regard, we explain optimal hyperparameter selection and training this model.

\begin{figure}
    \centering
    \includegraphics[width=.75\linewidth]{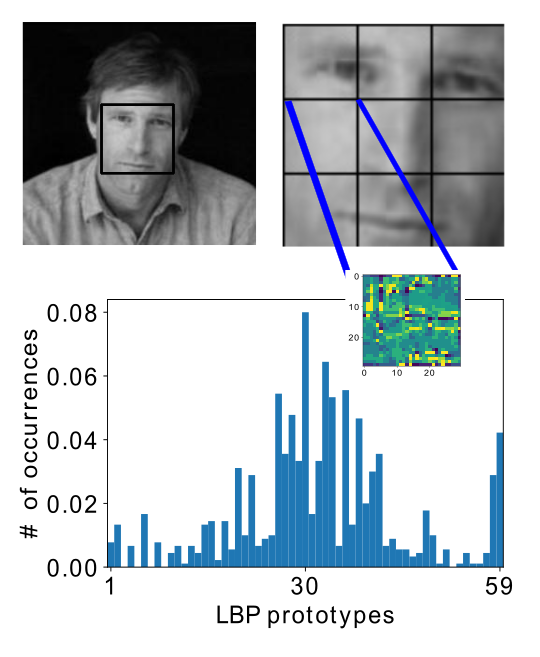}
    \hspace{40mm} (a)
    \includegraphics[width=\linewidth]{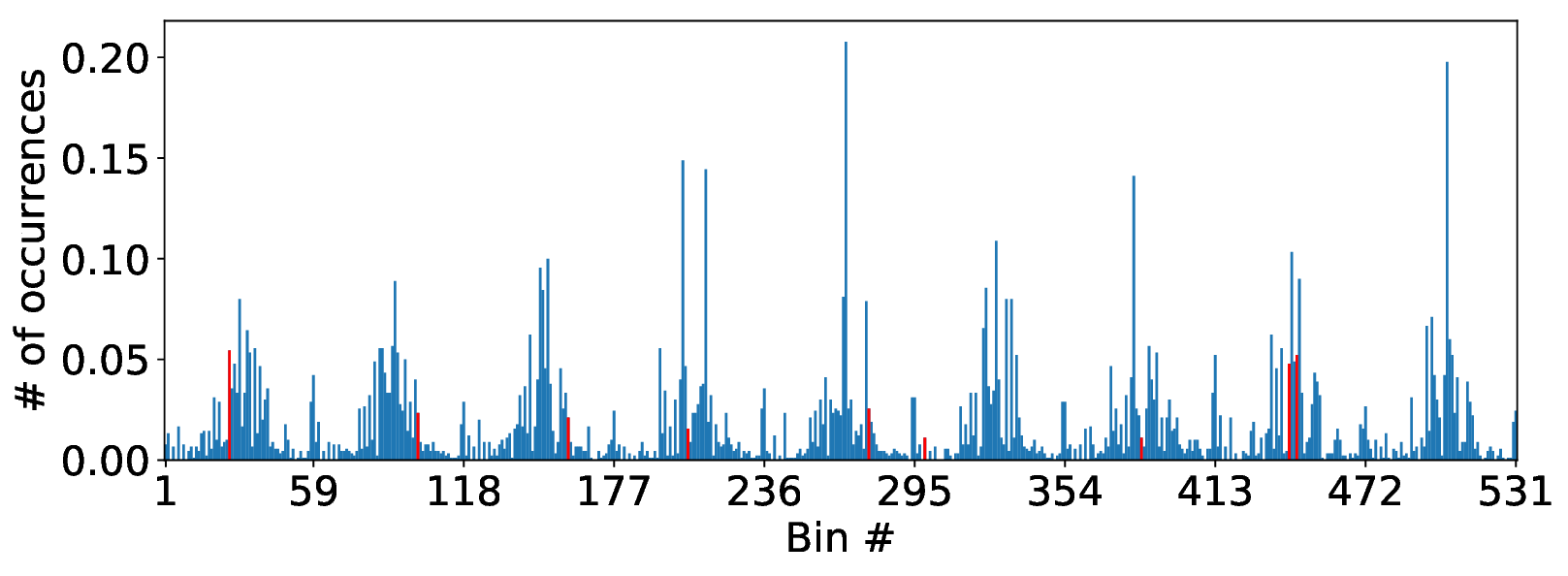}\\
    (b)
    \caption[Preprocessing]{
    Pictorial depiction of the preprocessing step.
    (a) Top two figures highlight the essential components of normalisation: first is the 2D grayscale image with the face detection box and second is the cropped and resized image.
    Top second figure and bottom figure together depict the essentials of feature extraction: we show $3\times3$ image segments, along with non rotation-invariant uniform LBP for one segment, and the 59-dimensional normalised 
    histogram representation for this LBP. 
    (b) Feature extraction and selection: spatially-enhanced histogram, generated by concatenating the nine LBP histograms, represents a 531-dimensional feature vector. The sparse red bars represent the low-dimensional feature vector after feature selection.}
    \label{fig:preprocess}
\end{figure}

The raw dataset~$\mathcal{D}_\ell^\text{raw}$~\eqref{eq:rawdata_l} comprises 125 LFW images of varying facial-region size, orientation and illumination, and hence this dataset needs to be normalised before being used for training $\varepsilon$-SVR models.
To normalise these images according to \texttt{normalise}~\eqref{eq:normalise}, we first convert each truecolor face image into its gray-scale version using a pre-defined function~\cite{rgb2gray} of the \texttt{OpenCV} package.
Next we crop the facial region of each gray-scale image by extracting the sub-matrix corresponding to the coordinates (integer part) in the face box.
Finally, each cropped image is resized to~$90\cross90$ ($m_\text{r}=n_\text{r}=90$ in Eq.~\eqref{eq:normalise}), which is an approximate average of the dimensions of 125 images, using \texttt{OpenCV}'s \texttt{resize} function~\cite{resize}; see~\cref{fig:preprocess}(a).
Additionally, for each normalised image~$\bm{I}^\text{norm}$, the scaled coordinate~$s^{(\ell)}$ is calculated using
\begin{equation}
\label{eq:scale_l}
    \texttt{scale}_\ell: \Reals \to \Reals: s^{\text{raw}}_\ell \mapsto s^{(\ell)},
\end{equation}
from the actual coordinate $s^{\text{raw}}_\ell$ and image dimensions.

We now apply \texttt{extract}~\eqref{eq:extract} on each $90\cross90$ image~$\bm{I}^\text{norm}$ to construct the corresponding feature vector~$\bm{x}^\text{norm}$. 
To this end, we choose local binary patterns (LBP)~\cite{MVBP13} as our image descriptor because local descriptors are robust with respect to pose and illumination changes in images and are invariant to hyperparameter selection~\cite{AHP06}.
We divide~$\bm{I}^\text{norm}$ into a $3\times3$ grid of equal segments and calculate the LBP histogram for each segment using the LBP implementation of Python's \texttt{scikit-image}~\cite{lbp}; see \cref{fig:preprocess}(a).
Upon choosing a circular (8,1) neighbourhood and restricting LBPs to only non-rotation-invariant uniform patterns, the LBP histogram for each segment has 59 bins, where 58 bins hold frequencies of 58 uniform patterns and all non-uniform patterns are counted in the remaining bin~\cite{parekh21}.
We use the spatially-enhanced histogram representation~\cite{AHP06}, which is the concatenation of the nine LBP histograms corresponding to the nine segments, as a 531-dimensional feature vector~$\bm{x}^\text{norm}$.
In \cref{fig:preprocess}(b), we represent this spatially enhanced histogram as an extended histogram plot with 531 bins.

In the final sub-step of preprocessing, i.e., feature selection, we use the correlation-based feature selection~(CFS)~\cite{CWWS14} technique and $\mathcal{D}_{\ell,\text{model}}^\text{raw}$.
The correlations between the feature vectors~$\{\bm{x}_i^\text{norm}\}$~\eqref{eq:extract} and their corresponding outputs~$\{s_i^{(\ell)}\}$~\eqref{eq:scale_l} are quantified by their Pearson correlation coefficients, which are calculated using the pre-defined function \texttt{pearsonr} of Python's \texttt{scipy}~\cite{pearson}.
By this Pearson CFS technique, we then reduce a 531-dimensional vector~$\bm{x}^\text{norm}$ to a $F_\ell$-dimensional vector~$\bm{x}^{(\ell)}$ according to
\begin{equation}
\label{eq:select_l}
    \texttt{select}_\ell: \Reals^{531} \to \Reals^{F_\ell}: \bm{x}^\text{norm} \mapsto \bm{x}^{(\ell)},
\end{equation}
where $F_\ell < 10$.
This bound for $F_\ell$ is chosen to avoid over-fitting during model calibration, when each model is trained using 10 feature vectors.
Although this huge reduction in $F_\ell$ can lead to over-generalisation, it does not influence the comparison between our classical and hybrid models.
In~\cref{fig:preprocess}(b), we use red bars to represent the selected features, whose indices are then used for selecting features of test images.

By performing the three operations~(\ref{eq:normalise},\ref{eq:extract},\ref{eq:select_l}) on each image in $\mathcal{D}^\text{raw}_\ell$ and the scaling operation~\eqref{eq:scale_l} on the corresponding landmark coordinate, we derive the preprocessed dataset~$\mathcal{D}^{(\ell)}$ as an union of two disjoint datasets~\eqref{eq:ModelTest_l_raw}.
In this regard, we obtain 
\begin{align}
\label{eq:ModelTest_l}
    \mathcal{D}^{(\ell)}&=\left\{\left.\left(\bm{x}^{(\ell)}_i,s^{(\ell)}_i\right) \right\vert i\in [N]\right\} \subset \Reals^{F_\ell} \times \Reals \nonumber \\
    &=\mathcal{D}_\text{model}^{(\ell)}\sqcup \mathcal{D}_\text{test}^{(\ell)}.
\end{align}
We use the dataset~$\mathcal{D}_\text{model}^{(\ell)}$ to construct an $\varepsilon$-SVR model
\begin{equation}
\label{eq:approxdetect_l}
    \widehat{\texttt{detect}}_\ell: \Reals^{F_\ell} \to \Reals: \bm{x}^{(\ell)} \mapsto \tilde s^{(\ell)},
\end{equation}
which approximately predicts the value of~$s^{(\ell)}$~\eqref{eq:scale_l} as~$\tilde s^{(\ell)}$. 
In order to compare classical vs hybrid algorithms, we generate three different~$\widehat{\texttt{detect}}_\ell$ models, namely SKL-SVR, SA-SVR and QA-SVR.
We fix the error tolerance as $\varepsilon=0.1$, which is the default value in \texttt{LIBSVM}~\cite{CL11} and its implementation in \texttt{scikit-learn}~\cite{sklearnSVR}, for all these three FLD models.
For the SA solver, we fix both the number of sweeps and the number of repetitions to 1000, and use the iteratively-averaged value over 20 low-energy samples as the solution for the QUBO problem~\cite{LFRL18}.
For the Hybrid Solver, we fix the parameter $\texttt{time\_limit}$ to 3s and 4s for the calibration and training steps, respectively, as our largest QUBO problem has 1000 variables~\cite{mintimelimit}.
Furthermore, we report the average prediction over 20 models as the value of~$\tilde s^{(\ell)}$ to account for the probabilistic nature of the SA and Hybrid Solvers~\cite{WWDM20}.

Before training an $\varepsilon$-SVR model using $\mathcal{D}_\text{model}^{(\ell)}$, we calibrate the model by tuning its hyperparameters.
To do this, we perform a search for optimal hyperparameter values over a grid defined by domains of the different hyperparameters.
We restrict the domain of each hyperparameter to a small subset based on certain assumptions and our observations.

Our choices for hyperparameter domains are now elaborated.
We fix $B_\text{f}=0$~\eqref{eq:encoding}, which is justified because fractional part was not required for getting feasible solutions using SVMs~\cite{WWDM20}. 
Consequently, $\gamma$~\eqref{eq:gamma} can only be certain integer values.
We pick $\gamma \in \{15,31,63\}$, corresponding to $B\in\{4,5,6\}$.
These bounds are empirically justified by observing insignificant changes in QUBO solutions with~$\gamma$ being outside this range.
We estimate the default value for the Gaussian kernel parameter~\eqref{eq:default_eta} as~$\eta=238$ using feature dimension~$F_\ell=6$ (on average) and variance of training dataset as 0.0007 (on average).
Assuming $\eta=238$ as the upper bound, we choose $\eta\in\{4,4^2,4^3,4^4\}$, which exponentially covers the domain for $\eta$ and is enough for our problem.
Furthermore, for the Lagrange multiplier~$\lambda$, we empirically choose a feasible subset $\{1,5,10\}$ as its domain.
To summarise, for SKL-SVR, the domain for each hyperparameter in the tuple~$(\gamma,\eta)$ is 
\begin{align}
\label{eq:hyperparameters_set1}
    \gamma \in \{15,31,63\},\; 
    \eta \in \{4, 4^2, 4^3, 4^4\},
\end{align}
and for both SA-SVR and QA-SVR, the hyperparameter tuple is $(B,B_\text{f},\eta,\lambda)$, with 
\begin{align}
\label{eq:hyperparameters_set2}
    & B \in \{4, 5, 6\},\;
    B_\text{f}=0,\; \nonumber\\
    &\eta \in \{4, 4^2,4^3,4^4\},\;
    \lambda \in \{1,5,10\},
\end{align}
being their corresponding domains.

The calibration step works as follows.
For each point on the grid, where a point represents tuple $(\gamma,\eta)$ for SKL-SVR and $(B, B_\text{f}, \eta,\lambda)$ for SA-SVR and QA-SVR, we construct a model and test its performance.
By randomly sampling $10\%$ of $\mathcal{D}^{(\ell)}_\text{model}$~\eqref{eq:ModelTest_l}, without replacement, we first generate a dataset~$\mathcal{D}^{(\ell)}_\text{train}$.
Using this dataset, we then train a model~\eqref{eq:approxdetect_l} and test it on the remaining $90\%$ of $\mathcal{D}^{(\ell)}_\text{model}$ to evaluate a mean normalised error~(MNE), which is similar to MNDE~\eqref{eq:MNDE}.
We define MNE for each coordinate of each landmark as
\begin{equation}
   \text{MNE}^{(\ell)} := \frac{1}{V}\sum_{m=1}^{V} \frac{|s^{(\ell)}_m-\tilde s^{(\ell)}_m|}{d_\text{c}},
\end{equation}
where $V=90$ is size of this test dataset, and $d_\text c=m_\text r=90$ and $d_\text c=n_\text r=90$ for $\ell$ representing $x$ and $y$ coordinate, respectively.
We re-run this procedure 50 times to calculate an average $\text{MNE}^{(\ell)}$ corresponding to each hyperparameter tuple.
After repeating this calculation for all points on the grid, we pick the tuple yielding the minimum value for average $\text{MNE}^{(\ell)}$.

We construct the final $\varepsilon$-SVR model~$\widehat{\texttt{detect}}_\ell$~\eqref{eq:approxdetect_l} using the whole dataset~$\mathcal{D}^{(\ell)}_\text{model}$ and the best hyperparameter tuple obtained from the calibration step.
The sub-task in Eq.~\eqref{eq:shape_l} is then approximately accomplished by following the composition
\begin{equation}
\label{eq:approxshape_l}
    \widehat{\texttt{shape}}_\ell= \texttt{rescale}_\ell\circ \widehat{\texttt{detect}}_\ell\circ \texttt{preprocess}_\ell,
\end{equation}
where $\texttt{rescale}_\ell$ is the inverse of $\texttt{scale}_\ell$~\eqref{eq:scale_l}.
The function~$ \widehat{\texttt{shape}}_\ell$ yields an approximate prediction~$\tilde s^\text{raw}_\ell$ of one coordinate of one landmark for each image in~$\mathcal{D}^{(\ell)}_\text{test}$. 
This prediction, along with the prediction for the other coordinate~(\cref{fig:MLworkflow}), are then used to construct and assess the FLD model~\eqref{eq:detect_landmark}.

\begin{widetext}
\section{QUBO formulation for SVR}
\label{appendix:qubo}
In this appendix, we establish an expression for elements of the QUBO matrix~$\widetilde{\bm{Q}}$~\eqref{eq:quboproblem} in QUBO formulation of a SVR.
To make the established expression symmetric, we treat the squared-penalty term in the objective function in~\cref{eq:minimize_unconstrained1} as a product of two similar terms.
Expanding the objective function~$\mathcal{L}(\bm{\alpha})$ yields
\begin{align}
\label{eq:minimize_unconstrained2}
\mathcal{L}(\bm{\alpha})  = & \frac12\sum_{n,m=0}^{2M-1}\alpha_n Q_{nm}\alpha_m + \sum_{m=0}^{2M-1} \alpha_m c_m+\lambda\left(\sum_{m=0}^{M-1}\alpha_m\right)^2 +\lambda\left(\sum_{m=M}^{2M-1}\alpha_m\right)^2\nonumber\\
&-\lambda\sum_{m=0}^{M-1}\sum_{n=M}^{2M-1}\alpha_m\alpha_n
-\lambda\sum_{m=M}^{2M-1}\sum_{n=0}^{M-1}\alpha_m\alpha_n. 
\end{align}
To obtain a binary form, we substitute the real-to-binary encoding in~\cref{eq:encoding} into this objective function,
and expand the quadratic terms.
We then obtain
\begin{align}
\label{eq:revObject}
\mathcal{L}(\bm{a})=
& \frac{1}{2^{2B_\text{f}+1}}\sum_{n,m=0}^{2M-1}
\sum_{i,j=0}^{B-1}
2^{i+j} a_{Bn+i} Q_{nm} a_{Bm+j}+ \frac{1}{2^{B_\text{f}}}
\sum_{n=0}^{2M-1}\sum_{i=0}^{B-1}
2^i c_n a_{Bn+i}+\frac{\lambda}{2^{2B_\text{f}}} \sum_{n,m=0}^{M-1}\sum_{i,j=0}^{B-1}2^{i+j} a_{Bn+i}a_{Bm+j} \nonumber\\
&+\frac{\lambda}{2^{2B_\text{f}}} \sum_{n,m=M}^{2M-1}\sum_{i,j=0}^{B-1}2^{i+j} a_{Bn+i}a_{Bm+j}- \frac{\lambda}{2^{2B_\text{f}}}
\sum_{m=0}^{M-1}\sum_{n=M}^{2M-1}
\sum_{i,j=0}^{B-1}2^{i+j} a_{Bn+i} a_{Bm+j}\nonumber\\
&- \frac{\lambda}{2^{2B_\text{f}}}
\sum_{m=M}^{2M-1}\sum_{n=0}^{M-1}
\sum_{i,j=0}^{B-1}2^{i+j} a_{Bn+i} a_{Bm+j}.
\end{align}
To fit the QUBO form, we express this equation as
\begin{equation}
    \mathcal{L}(\bm{a})
    = \bm{a}^\T \widetilde{\bm{Q}}\bm{a} =\sum_{n,m=0}^{2M-1}\sum_{i,j=0}^{B-1} a_{Bn+i}\widetilde{Q}_{Bn+i,Bm+j} a_{Bm+j},
\end{equation}
where $\widetilde{\bm{Q}}$ is the QUBO matrix with size~$2MB\times 2MB$.
Then by Eqs.~\eqref{eq:revObject},~\eqref{eq:Heaviside} and~\eqref{eq:flipHeaviside}, we have
\begin{align}
    \widetilde{Q}_{Bn+i,Bm+j}
    = &
    \frac12\frac{2^{i+j}}{2^{2B_\text{f}}} Q_{nm}
    + \frac{2^{i}}{2^{B_\text{f}}}
    \updelta_{nm}\updelta_{ij} c_n +\lambda\frac{2^{i+j}}{2^{2B_\text{f}}}
    \bar{\Theta}(n-M)\bar{\Theta}(m-M)+\lambda\frac{2^{i+j}}{2^{2B_\text{f}}}
    \Theta(n-M)\Theta(m-M)\nonumber\\
    &- \lambda\frac{2^{i+j}}{2^{2B_\text{f}}} \bar{\Theta}(m-M)\Theta(n-M)- \lambda\frac{2^{i+j}}{2^{2B_\text{f}}}
    \bar{\Theta}(n-M)\Theta(m-M).
\end{align}
The combination of this equation and the identity $\bar{\Theta}(i)\bar{\Theta}(j)+
    \Theta(i)\Theta(j) = 1-\bar{\Theta}(i)\Theta(j)-\bar{\Theta}(j)\Theta(i)$,
for any~$i,j\in \Integers$, yields the following expression for elements of the QUBO matrix:
\begin{align}
    \widetilde{Q}_{Bn+i,Bm+j}
    = &
    \frac12\frac{2^{i+j}}{2^{2B_\text{f}}} Q_{nm}
    + \frac{2^{i}}{2^{B_\text{f}}}
    \updelta_{nm}\updelta_{ij} c_n
    +\lambda\frac{2^{i+j}}{2^{2B_\text{f}}}- 2\lambda\frac{2^{i+j}}{2^{2B_\text{f}}} \bar{\Theta}(m-M)\Theta(n-M)- 2\lambda\frac{2^{i+j}}{2^{2B_\text{f}}}
    \bar{\Theta}(n-M)\Theta(m-M).
\end{align}

\section{Detailed results}
\label{app:results}
Here we present detailed numerical results of our experiments.
We state the optimal hyperparameter tuples in Eqs.~\eqref{eq:hyperparameters_set1} and~\eqref{eq:hyperparameters_set2} for the three $\varepsilon$-SVR models, namely SKL-\texttt{landmark}, SA-\texttt{landmark} and QA-\texttt{landmark}.  
Additionally, we provide the exact numerical values used to make the plots in~\S\ref{subsec:performance_comparison} for performance comparison.
\begin{table}[h!]
\begin{tabular}{|c|c||c|c|c|}
\hline
Landmark \# &  & SKL & SA & QA \\
& Coordinate & ~~~($\gamma$,$\eta$)~~~ & ~~($B$, $\eta$, $\xi$)~~ & ~~($B$, $\eta$, $\xi$)~~\\
\hline 
\hline
1 & x & (31,16) & (5,16,1) & (4,16,1)\\
 & y & (15,16) & (6,4,10) & (4,16,10)\\
\hline 
2 & x & (63,4) & (6,4,5) & (5,16,10)\\
 & y & (31,64) & (4,64,10) & (5,16,10)\\
\hline 
3 & x & (63,16) & (6,64,10) & (4,64,5)\\
 & y & (15,64) & (5,16,1) & (4,64,5)\\
\hline 
4 & x & (63,64) & (5,256,10) & (4,256,5)\\
 & y & (63,4) & (4,16,5) & (4,41)\\
\hline 
5 & x & (15,64) & (5,64,5) & (5,16,10)\\
 & y & (63,4) & (5,16,5) & (4,64,10)\\
\hline 
\end{tabular}
\caption{Optimal hyperparameter tuples, with $\varepsilon=0.1$ and $B_\text f=0$)}
\label{tab:optimal}
\end{table}

\begin{table}[]
\centering
\begin{tabular}{|c|c||c|c|c|c|c|c|c|c|c|c|c|c|}
\hline
\multicolumn{2}{|c|}{Model}  & \multicolumn{2}{|c|}{Landmark \#1} & \multicolumn{2}{|c|}{Landmark \#2} & \multicolumn{2}{|c|}{Landmark \#3} &
\multicolumn{2}{|c|}{Landmark \#4} &
\multicolumn{2}{|c|}{Landmark \#5}\\
\hline 
Type & Fold \# & MNDE & FR & MNDE & FR & MNDE & FR & MNDE & FR & MNDE & FR \\
\hline
\hline
 & 1 & 4.17 & 0 & 4.37 & 8 & 6.64 & 24 & 6.61 & 24 & 8.83 & 40\\
 & 2 & 5.07 & 8 & 3.45 & 4 & 5.72 & 12 & 7.13 & 24 & 6.55 & 24\\
SKL & 3 & 3.72 & 0 & 3.45 & 4 & 5.87 & 8 &6.38 & 8 & 5.97 & 4\\
 & 4 & 4.91 & 4 & 4.71 & 4 & 6.69 & 24 & 6.87 & 12 & 6.22 & 12\\
 & 5 & 4.19 & 0 & 4.14 & 4 & 6.03 & 16 & 5.42 & 12 & 6.28 & 12\\
\cline{2-12}
 & Mean & 4.41 & 2.4 & 4.02 & 4.8 & 6.19 & 16.8 & 6.48 & 16 & 6.77 & 18.4\\
 & & (0.0253) & & (0.0242) & & (0.0377) & & (0.0388) & & (0.0385) &\\
\hline
\hline
 & 1 & 4.53 & 4 & 4.35 & 8 & 7.41 & 20 & 7.21 & 20 & 8.19 & 32\\
 & 2 & 5.17 & 8 & 3.47 & 0 & 5.94 & 12 & 8.18 & 28 & 7.66 & 28\\
SA & 3 & 3.75 & 0 & 3.51 & 0 & 6.55 & 12 & 6.09 & 8 & 8.2 & 24\\
 & 4 & 4.95 & 8 & 4.84 & 4 & 7.13 & 24 & 7.93 & 28 & 6.73 & 16\\
 & 5 & 4.02 & 4 & 4.14 & 4 & 6.07 & 16 & 5.22 & 12 & 6.24 & 12\\
\cline{2-12}
 & Mean & 4.48 & 4.8 & 4.06 & 3.2 & 6.62 & 16.8 & 6.93 & 19.2 & 7.41 & 22.4\\
 & & (0.0249) & & (0.0235) & & (0.0375) & & (0.0403) & & (0.0407) &\\
\hline
\hline
 & 1 & 5.45 & 4 & 4.39 & 8 & 6.76 & 20 & 9.68 & 36 & 7.73 & 32\\
 & 2 & 5.36 & 8 & 3.72 & 0 & 5.95 & 16 & 11.77 & 56 & 7.71 & 24\\
QA & 3 & 4.02 & 4 & 3.38 & 0 & 6.31 & 20 & 9.40 & 44 & 7.23 & 16\\
 & 4 & 4.95 & 8 & 4.85 & 4 & 7.05 & 28 & 10.48 & 52 & 6.36 & 20\\
 & 5 & 3.88 & 0 & 4.18 & 4 & 5.95  & 8 & 5.21 & 12 & 6.29 & 16\\
\cline{2-12}
 & Mean & 4.73 & 4.8 & 4.10 & 3.2 & 6.41 & 18.4 & 9.31 & 40 & 7.07 & 21.6\\
 & & (0.0251) & & (0.0236) & & (0.0379) & & (0.0463) & & (0.039) &\\
\hline
\hline
\end{tabular}
\caption{Detailed results on 5-fold cross validation over 125 LFW images.
For each model, we state MNDE in \% and FR in \% and the standard deviation of normalised detection errors is shown in parentheses.
These values are used in~\cref{fig:compare_1}(a)}
\label{tab:k-fold}
\end{table}

\begin{table}[]
\centering
\begin{tabular}{|c|c||c|c|c|c|c|c|c|c|c|c|c|c|}
\hline
\multicolumn{2}{|c|}{Model}  & \multicolumn{2}{|c|}{Landmark \#1} & \multicolumn{2}{|c|}{Landmark \#2} & \multicolumn{2}{|c|}{Landmark \#3} &
\multicolumn{2}{|c|}{Landmark \#4} &
\multicolumn{2}{|c|}{Landmark \#5}\\
\hline 
Type & Fold \# & MNDE & FR & MNDE & FR & MNDE & FR & MNDE & FR & MNDE & FR \\
\hline
\hline
 & 1 & 4.81 & 4.27 & 5.2 & 10.37 & 7.05 & 23.17 & 6.86 & 20.12 & 7.51 & 25.61\\
 & 2 & 5.21 & 7.32 & 4.74 & 6.71 & 7.74 & 26.83 & 7.62 & 24.39 & 7.42 & 20.73 \\
SKL & 3 & 5.49 & 6.71 & 4.53 & 5.49 & 7.65 & 26.22 & 7.30 & 21.95 & 7.57 & 20.73\\
 & 4 & 4.90 & 5.49 & 5.09 & 7.93 & 6.43 & 17.07 & 7.09 & 20.73 & 6.99 & 17.68\\
 & 5 & 5.10 & 4.27 & 4.53 & 7.93 & 7.21 & 23.17 & 6.78 & 18.90 & 7.00 & 18.29\\
  & & (0.0280) & & (0.0318) & & (0.0468) & & (0.0346) & & (0.0344) &\\
\hline
\hline
 & 1 & 4.72 & 4.27 & 5.07 & 8.54 & 8.51 & 31.71 & 6.48 & 14.02 & 12.11 & 63.41 \\
 & 2 & 5.33 & 7.32 & 4.86 & 7.93 & 8.13 & 31.10 & 7.60 & 26.22 & 9.13 & 39.63\\
SA & 3 & 5.41 & 6.71 & 4.58 & 4.88 & 7.87 & 22.56 & 6.65 & 16.46 & 10.28 & 50.61\\
 & 4 & 5.34 & 7.32 & 5.05 & 8.54 & 7.36 & 21.34 & 7.98 & 28.66 & 8.65 & 31.71\\
 & 5 & 5.15 & 3.66 & 4.67 & 7.32 & 6.92 & 22.56 & 6.85 & 20.12 & 7.11 & 16.46 \\
  & & (0.0279) & & (0.0315) & & (0.0437) & & (0.0354) & & (0.0336) &\\
\hline
\hline
 & 1 & 4.52 & 5.49 & 5.22 & 9.76 & 7.31 & 24.40 & 8.08 & 33.54 & 8.30 & 34.15\\
 & 2 & 5.55 & 8.54 & 5.06 & 8.54 & 8.41 & 34.76 & 10.67 & 53.05 & 8.45 & 31.10 \\
QA & 3 & 5.80 & 6.71 & 4.63 & 5.49 & 8.80 & 34.76 & 9.77 & 47.56 & 9.08 & 36.59\\
 & 4 & 5.54 & 7.93 & 5.06 & 7.93 & 7.18 & 22.56 & 10.11 & 48.78 & 7.66 & 24.39\\
 & 5 & 5.23 & 4.27 & 4.77 & 7.93 & 6.81 & 21.95 & 6.73 & 17.68 & 7.30 & 17.07\\
  & & (0.0284) & & (0.0318) & & (0.0449) & & (0.0340) & & (0.0343) &\\
\hline
\hline
\end{tabular}
\caption{Detailed results on testing the models, which are trained during 5-fold cross validation, on 164 LFPW images.
For Fold \#5, we state the standard deviation of normalised detection errors in parentheses.
The values for rows `Fold \#5' are used in~\cref{fig:compare_2}(a).
}
\label{tab:lfpw_results}
\end{table}

\begin{table}[]
\centering
\begin{tabular}{|c|c||c|c|c|c|c|c|c|c|c|c|c|c|}
\hline
\multicolumn{2}{|c|}{Model}  & \multicolumn{2}{|c|}{Landmark \#1} & \multicolumn{2}{|c|}{Landmark \#2} & \multicolumn{2}{|c|}{Landmark \#3} &
\multicolumn{2}{|c|}{Landmark \#4} &
\multicolumn{2}{|c|}{Landmark \#5}\\
\hline 
Type & Fold \# & MNDE & FR & MNDE & FR & MNDE & FR & MNDE & FR & MNDE & FR \\
\hline
\hline
 & 1 & 4.12 & 2.09 & 4.12 & 2.83 & 6.07 & 8.20 & 5.75 & 10.74 & 5.35 & 9.47\\
 & 2 & 4.65 & 3.73 & 4.26 & 2.24 & 8.00 & 29.53 & 5.99 & 11.71 & 5.05 & 6.41\\
SKL & 3 & 4.53 & 3.50 & 4.02 & 1.49 & 6.81 & 18.42 & 5.48 & 9.40 & 5.15 & 7.23\\
 & 4 & 4.17 & 3.50 & 4.48 & 2.61 & 5.82 & 9.32 & 5.46 & 10.29 & 4.90 & 6.04\\
 & 5 & 4.25 & 2.68 & 3.61 & 1.19 & 6.57 & 14.47 & 5.80 & 11.56 & 4.89 & 7.68\\
  & & (.0257) & & (.0220) & & (.0339) & & (.0335) & & (.0327) &\\
\hline
\hline
 & 1 & 3.88 & 1.94 & 4.30 & 3.06 & 5.77 & 7.08 & 5.60 & 10.96 & 8.42 & 29.68\\
 & 2 & 4.78 & 3.65 & 4.37 & 2.76 & 7.84 & 27.82 & 5.90 & 12.38 & 6.04 & 7.68\\
SA & 3 & 4.44 & 3.06 & 4.19 & 1.57 & 6.47 & 16.55 & 5.28 & 9.32 & 7.01 & 13.12 \\
 & 4 & 4.77 & 3.95 & 4.51 & 2.46 & 6.06 & 10.14 & 7.53 & 21.92 & 5.93 & 8.80\\
 & 5 & 4.33 & 2.61 & 3.82 & 1.72 & 6.22 & 12.83 & 5.64 & 10.66 & 4.83 & 7.23\\
 & & (.0259) & & (.0229) & & (.0346) & & (.0332) & & (.0323) &\\
\hline
\hline
 & 1 & 3.56 & 1.57 & 4.07 & 3.21 & 5.70 & 5.59 & 7.79 & 23.71 & 5.67 & 9.55 \\
 & 2 & 5.08 & 4.47 & 4.50 & 3.65 & 8.21 & 31.92 & 9.61 & 42.95 & 5.71 & 7.08\\
QA & 3 & 4.58 & 3.28 & 4.22 & 2.01 & 7.88 & 28.71 & 9.48 & 40.19 & 6.18 & 9.40\\
 & 4 & 5.00 & 4.55 & 4.46 & 2.24 & 5.47 & 6.86 & 10.29 & 50.71 & 5.43 & 8.13\\
 & 5 & 4.42 & 2.83 & 3.84 & 1.49 & 5.94 & 10.51 & 5.51 & 10.44 & 4.90 & 6.94\\
 & & (0.0261) & & (.0227) & & (.0328) & & (.0330) & & (.0323) &\\
\hline
\hline
\end{tabular}
\caption{Detailed results on testing the models, which are trained during 5-fold cross validation, on 1341 BioID images.
For Fold \#5, we state the standard deviation of normalised detection errors in parentheses.
The values for rows `Fold \#5' are used in~\cref{fig:compare_2}(b).}
\label{tab:bioid_results}
\end{table}

\end{widetext}

\end{document}